# Feather: A Feature Model Transformation Language

Ahmet Serkan Karataş [*]

**Abstract:** Feature modeling has been a very popular approach for variability management in software product lines. Building a feature model requires substantial domain expertise, however, even experts cannot foresee all future possibilities. Changing requirements can force a feature model to evolve in order to adapt to the new conditions. Feather is a language to describe model transformations that will evolve a feature model. This article presents the structure and foundations of Feather. First, the language elements, which consist of declarations to characterize the model to evolve and commands to manipulate its structure, are introduced. Then, semantics grounding in feature model properties are given for the commands in order to provide precise command definitions. Next, an interpreter that can realize the transformations described by the commands in a Feather script is presented. Finally, effectiveness of the language is discussed using two realistic examples, where one of the examples includes a system from a dynamic environment and the other employs a system that has a large feature model containing 1,227 features.

## 1. Introduction

The idea of using product lines has proven to be useful in various areas ranging from automotive industry to electronic component manufacturing for over a century. As software products and software intensive systems have become more complex, adaptation of product line methodology within the realm of software engineering has attracted more attention. The combined efforts of researchers and practitioners led the way to the software product lines (SPLs), which are defined as *"a set of software intensive systems that share a common, managed set of features satisfying the specific needs of a particular market segment or mission and that are developed from a common set of core assets in a prescribed way"* [16].

SPLs have been successfully used to develop a wide variety of software products and software intensive systems at lower costs, in shorter times, and with higher quality [40]. For instance, Salion Inc. used a reactive software product line engineering approach to develop revenue acquisition management systems [15]. The Boeing Company adopted the Bold Stroke SPL architecture to develop avionic software families [41].

Success of SPLs has encouraged researchers to widen the area of usage to include dynamic systems that can face fluctuating context conditions, changing user requirements, need to include and facilitate novel resources rapidly, and so on. The constant need for adaptation required to enhance the variability management capabilities of SPLs to cover runtime, which gave birth to the dynamic software product lines (DSPLs). In addition to the properties they inherit from SPLs, DSPLs have several other properties such as dynamic variability management, flexible variation points and binding times, context awareness, and self-adaptability [21].

A major factor affecting the success of a software product line, classic or dynamic, is the variability management activity it incorporates [12]. In the literature there are reports of various variability modeling and management approaches such as feature modeling (e.g., [28]), using UML and its extensions (e.g., [46]), and using domain ontologies (e.g., [4]). Feature modeling stands out as a very popular choice within a wide variety of techniques that can be used for this purpose [11].

If the famous proposition *"everything changes and nothing stays still"* attributed to the philosopher Heraclitus is true, then feature models are not exceptions. For instance, consider a scenario where a company offers a mediator platform between vendors that provide different web services and their customers. Since different customers can be interested in different types of services, company adopts a DSPL approach to create custom tailored products and uses feature models to model and manage variability. As the customer base for the platform grows, vendors that provide new services request to join the platform and the company decides to include some of them in the platform to expand the range of services provided. When complaints from users about some services increase, company decides to remove services that do not meet certain quality requirements from the platform. The former decision requires adding new variation points, variants, and constraints to the feature model, whereas the latter requires updating or removing some of the existing variation points, variants, and constraints from it.

Factors guiding the feature model design can change and necessitate a change in the feature model structure. It can be necessary to add new features, update some outdated decomposition relations, remove some of the existing constraints, etc., to meet the new requirements imposed by the changing conditions. Hence, changing conditions can force a feature model to evolve. However, evolution at runtime by manipulating the variability model and evolution of DSPLs remain to be an important challenge [22].

[*] Department of Computer Engineering, Middle East Technical University, Ankara, Turkey (e-mail: karatas@ceng.metu.edu.tr).

Industrial reports back up this view. A survey shows that feature modeling is the most frequently used approach for variability management in the industrial practice [7]. The same survey also indicates that the most frequently reported problem by the practitioners is the variability model evolution.

Feather is proposed to contribute to the solution of this problem. It is a feature model transformation language. It provides commands to add, update, and remove feature model elements (i.e., features, decomposition relations, and cross-tree constraints) for manipulating the structure of a feature model. These commands can be used to write a script that instructs how to transform a feature model to adapt to the changing requirements.

There exist several model transformation languages to process models conforming to various meta-models. However, model transformation is an intrinsically difficult task [13]. To specify even a simple transformation can require a lot of learning time and the steep learning curve can be an inhibitive factor in the usage of many model transformation frameworks [1]. Feather aims to provide simple constructs that are natural and intuitive for feature modelers.

The need for evolution can be in constant demand (e.g., when the feature model is used in a dynamic system) or require a lot of effort (e.g., when the model to evolve is very large), hence, support for automated transformation is desirable. Therefore, an interpreter that can realize the transformations described by Feather commands has been developed in order to provide automated support. It can read and validate an input Feather script, execute the specified commands and apply the described transformations to the input feature model, and save the resulting model in the specified format.

An important characteristic of Feather that is worth mentioning is the provided feature variable concept. A feature variable is a description for a feature or a group of features, which describes the features via their properties. The model elements that commands operate on are features or relations involving features. Modelers can facilitate feature variables to describe the feature model elements that will be involved in a command and leave the task of figuring out the actual features to the interpreter. This strategy shifts the feature identification task from coding time to runtime and enhances flexibility significantly. In addition to this, feature variables enable operating on feature models where only general guidelines on the structure of a feature model are available and the exact structure of the model is not known.

The remainder of this article is organized as follows. Section 2 provides a brief background to feature models and model transformation. Sections 3, 4, and 5 present Feather. Section 6 discusses semantics that is grounded in the properties of the input and the transformed feature models. Section 7 presents a Feather interpreter, components and the third party tools incorporated in, and the tasks performed by it. Section 8 presents an evaluation for the language via discussing two realistic examples: one from a system with a highly dynamic nature and one from a system that has a large feature model that contains 1,227 features. Section 9 discusses related work. Finally, Section 10 contains the conclusion.

## 2. Background

### 2.1 Feature Models

Modeling and managing variability in software product lines is a key activity. Since their introduction by Kang et al. [28] feature models have been widely used by both academic and industrial communities for this purpose [11]. A feature model essentially represents the set of valid configurations derivable from it and consists of three types of elements: features, decomposition relations, and cross-tree constraints.

A feature is a distinguishable characteristic of a concept that is relevant to a stakeholder of the concept [44]. Features can be used to represent different entities. For instance, Kang et al. [29] propose to use features to represent capability, operating environment, domain technology, or implementation technique related entities.

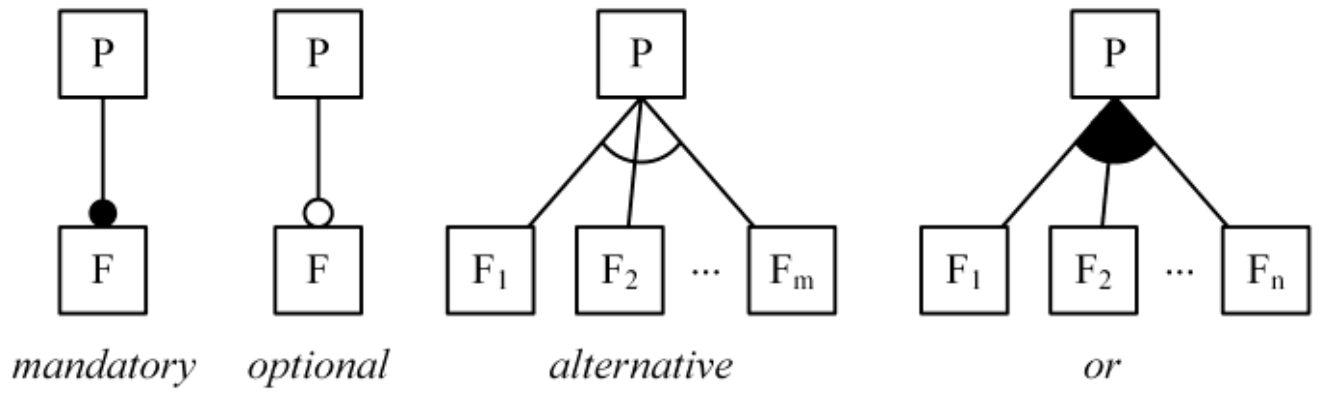

Fig. 1. Decomposition relation types

Decomposition relations describe the hierarchical structure of a feature model by representing the relations between a parent feature and its children. There are four types of decomposition relations: *mandatory*, *optional*, *alternative*, and *or*. If there is a mandatory relation between a parent and a child, then every configuration that includes the parent must also include the child. An optional relation between a parent and a child means that a configuration including the parent can include or exclude the child. If there is an alternative relation between a parent and a group of children, then a configuration that includes the parent must include exactly one child from the group. If there is an or relation between a parent and a group of children, then a configuration that includes the parent must include at least one child from the group.

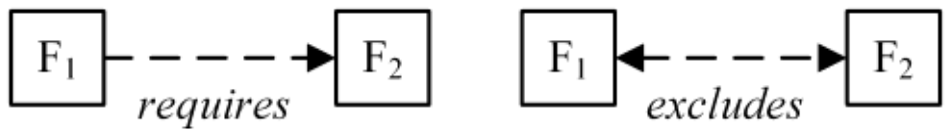

Fig. 2. Cross-tree constraint types

Cross-tree constraints define relations among features that can reside anywhere in the hierarchical structure. There are two types of cross-tree constraints: *requires* and *excludes*. If there is a requires constraint between two features, such as $F_1$ *requires* $F_2$, then a configuration that includes $F_1$ must also include $F_2$. An excludes constraint between two features $F_1$ and $F_2$ means that no configuration can include both of these features.

An important extension devised to feature models is the introduction of feature attributes [18]. A feature attribute can provide information about any measurable characteristic of the feature it belongs. For instance, attributes can be used to provide information about a feature such as cost of the feature, binding time for the feature, risk assessment for the feature, etc.

### 2.2 Model Transformation

A model transformation essentially generates a target model from a source model automatically [31]. It is performed in accordance with a transformation definition, which is a set of transformation rules. A transformation rule describes how one or more source model entities can be transformed into the elements in the target language to construct the target model. Reader can refer to the work by Mens and Van Gorp [36] for a detailed taxonomy on model transformation. Here, the elements relevant to this article are presented briefly.

A transformation can be horizontal, where the source and target models are at the same abstraction level, or vertical, where the abstraction levels are different (e.g., refinement of abstract model constructs into concrete implementation elements). Feather enables horizontal transformations.

A model transformation can have only syntactical effects (e.g., transforming an abstract program syntax into a concrete programming language code, where the semantics of the program is retained), or can have more complex semantical implications. Semantics for a feature model is given by the set of configurations it represents [19]. Changing the structure of a feature model by a transformation can affect the set of configurations represented by the model, thus, Feather transformations can, and probably will, have semantical effects.

A transformation is endogenous if the source and target models are expressed in the same language and exogenous otherwise. Feather performs endogenous transformations.

A transformation can include multiple source and/or target models (e.g., while merging two or more models into one), or only one model where the modifications are performed directly on that model. If there is only one model then the transformation is called in-place, otherwise out-place. Feather performs in-place transformations.

## 3. Feather Basics

### 3.1 The Meta-model

The feature models that will be transformed with Feather must comply with the Feather meta-model, which is a basic feature model extended with feature attributes. It does not include cardinality-based decompositions as proposed in [19] or complex cross-tree constraints as described in [30]. Meta-model hierarchy is in the tree-structure (e.g., models in the form of directed acyclic graphs do not comply).

A feature can include any number of attributes. In addition to the attributes declared by the modeler, every feature is assumed to include four structural attributes, one to represent the feature and three to represent the decomposition relation between the feature and its parent. The first structural attribute, `_name`, stores the name of the feature as a string. The second one, `_parent`, stores the name of its parent as a string. The third one, `_decomp`, denotes the decomposition relation type between the feature and its parent. Finally, the last one, `_decompID`, stores an integer value to identify the sibling features that figure in the same group decomposition relation (i.e., *alternative*, *or*) with the feature (i.e., all the other features that figure in the same group decomposition relation will have the same value and no other feature will have that value, in their respective attributes). The first three can be read and/or updated, whereas the last one is a read-only attribute from the modeler's point of view.

### 3.2 Data Types, Operators, and Expressions

Feather includes four general data types (i.e., integer, real, Boolean, and string) and one feature model data type (i.e., the decomposition relation data type that contains the values {`mandatory`, `optional`, `alternative`, `or`}). Expressions can involve literals from these data types as well as <feature–attribute> terms. A <feature–attribute> term is represented with the notation *"FN".attr* and interpreted as 'the value of the attribute *attr* of the feature with the name "*FN*"'.

There are two types of expressions in Feather: arithmetic and logical. An arithmetic expression can consist of a single arithmetic operand or applications of arithmetic operators (i.e., addition, subtraction, multiplication, division, modulo, and additive inverse) to arithmetic expressions. These operators act same as their counterparts in the major programming languages such as C, except for the division operator. Division operator is always interpreted in the mathematical sense (e.g., `7 / 2` will yield 3.5, not 3). This design decision is taken in order to increase the convenience for modelers who can be non-programmers.

Logical expressions can consist of a single Boolean operand, application of a relational operator (i.e., <, <=, >, >=, =, and <>) to two arithmetic expressions, application of an equality check operator (i.e., `=` and `<>`) to two compatible values (e.g., two strings), or applications of the logical operators (i.e., `and`, `or`, and `not`) to logical expressions.

The priority of the operators are the same as their counterparts in the C programming language. Parentheses can always be used to alter the order of execution. When a <feature–attribute> term figures in an operation, the value of the attribute it refers to must be compatible with the operator for the expression to be considered well formed.

### 3.3 Feature Variables

Consider a scenario where a software company releases a video game that involves many game elements such as levels, maps, and characters. Every separate game installation models and manages the relations among the game elements via a feature model of its own. There are many third-party game elements the gamers can obtain and attach to their systems. Hence, the feature models constantly change (e.g., new game character features are added) and can be different from gamer to gamer. Company wants to release an update that will replace some of the third-party game characters (e.g., those that do not possess certain quality characteristics) with company issued characters in the systems of gamers. Since the game has a large gamer base, analyzing the feature model of every individual gamer and identifying the attached non-fitting features to be replaced is not a feasible strategy due to the enormous time and effort it requires. A smart solution is necessary, which can be formulated even if the DSPL modeler that prepares the update does not know which actual features will be affected.

Consider another scenario where a company manages a web service platform that offers numerous web services ranging from finance to fashion. Platform structure is modeled as a feature model and maintained in the platform server. Any third party service provider complying with the requirements can register her service to the platform. It is a very popular platform where thousands of services have already been registered, thus, the feature model includes thousands of features. Company decides to restructure the feature model to reflect the advances in the web technology (e.g., moving services that have been coded with pre-HTML 5 versions to a specific branch). Identifying the features that will be affected one-by-one and carrying out the restructuring task manually is extremely time consuming and error-prone due to the large number of features that must be checked. A smart solution that can automate the task and increase efficiency is necessary.

Feather provides feature variables to devise smart solutions for cases like the ones mentioned in the above scenarios. Feature variables enable delaying the feature identification process until the realization time and shift the workload needed for the identification from the modeler to the computer. They are used (by the modeler) to stand for a feature or a set of features with certain properties while describing a transformation step, and resolved (by the computer) into an actual feature or a set of actual features that has the desired properties while realizing the transformation step. For instance, the modeler in the first scenario can utilize a feature variable to have the effect *"let FV be a feature variable that represents all features such that; feature corresponds to a game character, the character belongs to the sorcerer class, the character does not have a multiplayer mod, and the character has a low resolution skin. Remove all features represented by FV from the feature model"* in the update script.

Scope of a feature variable is limited to the command it is used in. A feature variable describes a feature in a command through the attributes of it. For instance, if the term *F.attr* is used, then the feature variable *F* can resolve into a feature that has an attribute with the identifier *attr*. <Feature variable–attribute> terms can also figure in expressions. For instance, a condition such as *"value of F.attr must be greater than 50"* can be defined on a feature variable. In that case, the possible resolutions for *F* will be limited to features whose attributes satisfy all such conditions.

However, the aforementioned resolution strategy can lead to complications in some rare cases. For instance, assume that a modeler wants to specify a constraint conditionally, such as *"the feature represented by this variable must be priced less than 10 USD or must not require an internet connection"*. A feature satisfying one of these conditions can satisfy the modeler's needs. Such conditions can be expressed as *(V.price < 10 or V.reqintcon = false)*.

Assume that there is a feature *"F1"* that has a price tag 5 USD, however, has nothing to do with internet connection, hence, does not include such an attribute. Obviously, this feature can satisfy the programmer, however, it will be eliminated from the set of the variable's candidate resolutions since it does not have an attribute named *reqintcon*, and consequently, will not be included in a solution set. Such problems can easily be handled by introducing extra feature variables. For instance, this case can be expressed as *(V1.price < 10 or V2.reqintcon = false) and (V._name = V1._name or V._name = V2._name)*.

## 4. Feather Declarations

First part of every Feather script consists of declarations that define the model that will be transformed. There are three types of elements that must be provided to define a feature model: *(i)* the features contained in the model, *(ii)* the decomposition relations among the features in the model hierarchy, and *(iii)* the cross-tree constraints among features. Feather provides two declaration structures to describe features, one for the root feature and one for the non-root features, and one declaration structure to describe the cross-tree constraints. Information needed to infer the model hierarchy (i.e., decomposition relations) is included in the feature declarations.

### 4.1 Root Feature Declaration

Declarations part always starts with the declaration of the root feature (i.e., the feature model cannot be empty). Since a root feature does not have a parent, and consequently a decomposition relation, declaration for it does not include compartments for these information. A root feature declaration has the following form:

`root` *Feature-Name* [*Attribute-Declarations*]

*Feature-Name* is a string that provides the name of the root feature. *Attribute-Declarations* provide the <attribute–value> pairs to be included in the feature. A root feature declaration can involve any number of <attribute–value> pairs. An attribute declaration has the following form:

`attribute` *attribute-identifier value*

*attribute-identifier* provides the identifier of the attribute and complies with the rules for naming identifiers in the C programming language, except for it must start with a lowercase letter. *value* is a literal that can be chosen from the types integer, real, Boolean, or string.

For instance, the following root feature declaration includes two attributes: *price* and *core*, with the values *14.75* and *true*, respectively.

```
root "Root Name"
  attribute price 14.75
  attribute core  true;
```

### 4.2 Non-root Feature Declaration

The root feature declaration can be followed by any number of non-root feature declarations. A non-root feature declaration provides the name, decomposition relation information, and attribute declarations for a feature, and has the following form:

```
feature Feature-Name
    Parent-Name Decomposition-Relation-Info
    [Attribute-Declarations] ;
```

*Parent-Name* is a string that provides the name of the parent the feature will be connected to as a child. *Decomposition-Relation-Information* provides the type of the decomposition relation, which can be `mandatory`, `optional`, `alternative`, and `or`. For instance, the following code portion declares a feature that will be a child of the feature *"Parent"* and join to the same `or` decomposition relation with the feature *"Sibling"*.

```
feature "Feature Name"
  "Parent" or to "Sibling"
  attribute level 5
  attribute price 21.2
  attribute stype "Cat-A";
```

Every non-root feature in the feature model must be declared exactly once (i.e., duplicate declarations are not allowed and an existing feature's declaration cannot be omitted), however, ordering of the declarations is not important (e.g., a child feature can be declared before its parent). Declarations must comply with the tree-structure.

### 4.3 Cross-tree Constraint Declaration

Cross-tree constraint declarations constitute the final part of the declarations in a Feather script and can include any number of declarations. A cross-tree constraint declaration has the following form:

```
constraint Feature-1 Constraint-Type Feature-2
```

Features (i.e., *Feature-1* and *Feature-2*) included in the cross-tree constraint declaration must be declared before. *Constraint-Type* denotes the type of the constraint and can be `requires` or `excludes`.

```
constraint "Chat" requires "Internet" ;
```

Ordering of the cross-tree constraint declarations is not important. Although it is not recommended, since it is not good feature model design practice, duplicate declarations or declarations of the same effect (e.g., including both of the declarations *"X" excludes "Y"* and *"Y" excludes "X"*) are allowed. Such redundancies are eliminated while constructing the model.

## 5. Feather Commands

A feature model transformation consists of a combination of three basic activities: feature manipulation, decomposition relation manipulation, and cross-tree constraint manipulation. Feather provides ten command types to describe these manipulations. These command types are discussed in detail in the following subsections.

A command describes what must be done in order to carry out a transformation, not how to do it, hence, Feather adopts a declarative style. Every command consists of four compartments. The first compartment includes a command identifier that denotes the type of the command. The second one describes the entity or the set of entities that will be subject to the command. The third compartment provides additional information needed to execute the command. The fourth one describes the condition that must be satisfied for the command to take effect; if it cannot be satisfied then the command will have no effect on the model. The third and the fourth compartments are optional and can be omitted if not needed.

Consider a hypothetical case where a company, Company X, offers a family of web-based services/applications. Company X manages the service packages via the feature model depicted in Fig. 3. This hypothetical company and the simple feature model it employs will be used to present the Feather commands in the following subsections.

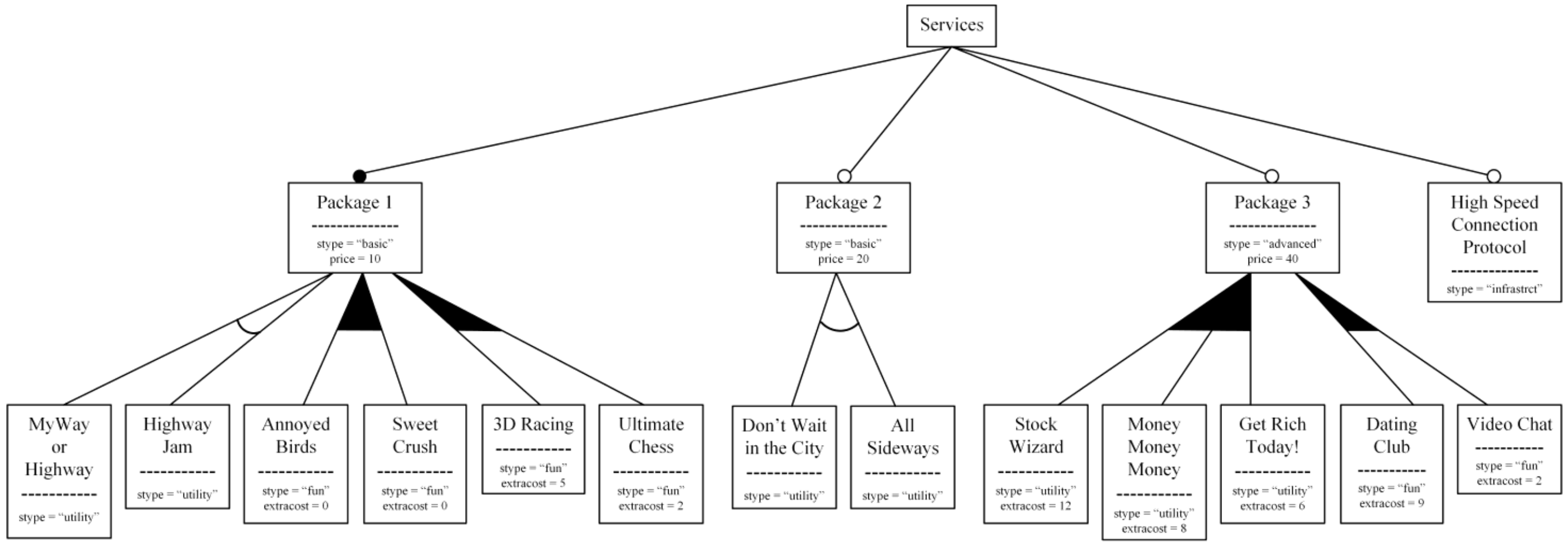

Fig. 3. The services feature model

### 5.1 add feature

Feature models can grow by the addition of new features and Feather provides the `add feature` command to meet these needs. Information required for adding the feature successfully consists of three groups: *(i)* name of the feature, which must be provided as a string, *(ii)* specification of the decomposition relation the added feature will figure in as a child, which must be specified via assignments to the structural attributes, and *(iii)* domain-related information about the feature, which must be provided via attribute declarations. Modeler can employ any number of feature variables while providing this information. The `add feature` command has the following form:

```
add feature Feature-Name
  with attributes ( Attribute-List )
  [Where-Clause] ;
```

*Feature-Name* is a string that provides the name of the feature that will be added. *Attribute-List* is a comma-separated list that includes the assignments to the structural attributes `_parent` and `_decomp`, which will provide the location and decomposition relation information for the feature that will be added, and other attribute assignments if there are any. The *Where-Clause*, which is an optional compartment, is a Boolean expression. An attribute assignment has the following form:

```
attribute-identifier = attribute-type : value
```

*attribute-type* can be one of the following: `numeric` (i.e., integer or real), `boolean`, `string`, and `inherited` (i.e., attribute value is inherited from another feature attribute and the type of the attribute will be the same type with the attribute that provides the value). *value* can be an expression of the specified type.

***Scenario***: Company X decides to add a bridge game to the set of provided applications in order to appeal to a larger user base, hence, the feature model must grow. Since the feature model is organized according to some guidelines determined by the marketing department, the new feature must be added to a location that will conform to these guidelines and have a number of attributes to describe the properties of the feature. ∎

The following command can be used to add the new feature representing the bridge game added to the system:

```
add feature "Bridge Pro"
  with attributes (
    _parent = P._name,
    _decomp = or to ASibling,
    stype   = string : "fun",
    extracost = numeric : 8)
  where P.stype = "basic"
   and  P.price <= 15
   and ASibling._parent = P._name
   and ASibling.extracost > 0;
```

The feature to be added (i.e., *"Bridge Pro"*) will be inserted as a child of the feature represented by the feature variable *P*, which must be resolved into an actual feature while realization. It will be in an `or` decomposition relation with the feature represented by the variable *ASibling*. It will have two attributes: *stype* and *extracost*. The condition to guide the resolutions is specified in the `where` compartment. For instance, *P* will resolve into a feature that has at least two attributes; *stype* and *price*, where the value of the former is *"basic"* and the value of the latter is less than or equal to *15*. When executed, this command will add the new feature as a child of the feature *"Package 1"*, in the same or decomposition relation with the features *"3D Racing"* and *"Ultimate Chess"*.

A feature addition command must be unambiguous on the specification of the parent, the decomposition relation that will connect the added feature to its parent, and the values of the attributes that will be included in the feature. If one of these presents an ambiguity then the command will have no effect on the feature model. For instance, if the *P.price <= 15* part is omitted from the `where` clause then the parent becomes ambiguous, since *P* can resolve into both *"Package 1"* and *"Package 2"*, and the feature will not be added to the model.

A key point in understanding ambiguities is noting that they are tightly related to feature variable resolutions, however, multiple valid resolutions do not necessarily lead to ambiguities. For instance, the variable *ASibling* will resolve into two features during the execution of the above command: *"3D Racing"* and *"Ultimate Chess"*, since both satisfy all the criteria specified for *ASibling*. However, both resolutions will lead to the same conclusion about the decomposition relation the new feature will join. Hence, there is no ambiguity and the command will successfully add the new feature.

The transformation described by a feature addition command must also be consistent with the feature model. For instance, a feature addition command must not try to insert a feature under *"Package 1"* and put it in the same alternative relation with *"Don't Wait in the City"*, which has a different parent. If the described transformation does not fit in the model structure, then the command will have no effect.

### 5.2 update feature

Properties of an entity represented by a feature can change in time, necessitating a feature update. Feather provides the `update feature` command type to meet these needs. Modeler must specify the feature to be updated and its attributes that will receive new values. An `update feature` command can be used to move a feature, change the decomposition relation type that connects a feature to its parent, or update the value of the attribute(s) of a feature. It has the following form:

```
update feature Feature-Descriptor
  set Attribute-Updates
  [Where-Clause] ;
```

*Feature-Descriptor* can be a string that includes the name of the feature that will be updated, or a feature variable that will describe the feature to be updated. *Attribute-Updates*, which is a comma-separated list, includes the updated attribute assignments. The *Where-Clause*, which is an optional compartment, is a Boolean expression.

***Scenario***: When the monthly sales reports arrive, company executives see that the sales for the *"Dating Club"* service have fallen dramatically. A quick analysis reveals that users find the service useful, however, too expensive. Company decides to lower the extra cost demanded from the users for that service, consequently, the feature representing the service must be updated. ■

The following command can be used to perform the task discussed in the above scenario:

```
update feature "Dating Club"
  set extracost = numeric: 5;
```

Modeler can employ as many feature variables as she requires while describing the feature to be updated or the new values that the attributes will receive. Value of any attribute, including the structural attributes (except for `_decompID`), of a feature can be updated. However, a feature update command must not be ambiguous about the new value an attribute will receive, otherwise, it will have no effect on the model. Similarly, if an update command is ambiguous about the feature to be updated (i.e., a feature variable is used to describe the feature to be updated and it can resolve into two or more features) or violates feature model design rules (e.g., try to change the name of the feature into a value that is already in use by another feature), then it will have no effect on the model.

***Scenario***: Marketing personnel feel that *"Package 2"* can become more attractive to the potential users if it includes fun related services. Therefore, they ask one of the fun services to be moved from *"Package 3"* to *"Package 2"*. ■

A feature can be moved to a different location in the model by updating the value of its structural attribute *_parent*. The type of the decomposition relation a feature has with its parent can be changed by updating the value of its structural attribute *_decomp*. The structural attribute *_decompID* is a read-only attribute from the perspective of a modeler, whose value can be retrieved, however, not updated. For instance, the following update command can be used to move the feature *"Dating Club"* to the desired location in the above scenario:

```
update feature "Dating Club"
  set _parent = "Package 2",
      _decomp = optional,
      extracost = numeric: 8;
```

Feature variables can be used to describe the feature to be moved, the new parent, and the new decomposition relation type. A decomposition relation update command can be used to update the values of a feature's attributes while manipulating the decomposition relation it figures in as a child feature as the above example shows. An update command must be unambiguous about which feature's decomposition relation will be updated, what the new parent will be, and the specifics about the new decomposition relation, otherwise, the command will have no effect on the model. For example, the following command will have no effect on the model since it is ambiguous about which feature (i.e., *"Dating Club"* or *"Video Chat"*) will be moved.

```
update feature F
  set _parent = "Package 2",
      _decomp = optional
  where F._parent = "Package 3"
   and  F.stype   = "fun";
```

Moving a feature implies moving the entire subtree descending from the moved feature. For instance, it the feature *"Package 1"* is moved, then all the features descending from it (*"My Way or Highway"*, *"Annoyed Birds"*, etc.) will be moved along with it. A decomposition relation update command cannot violate the tree-structure hierarchy of the feature model (e.g., cannot lead to a cycle in the model hierarchy), otherwise, it will have no effect.

An update command cannot be used to introduce new attributes to a feature or delete existing attributes from it. However, it is still possible to get this effect via an indirect strategy: a new feature with the desired set of attributes can be added, the children of the old feature can be moved to the new feature, cross-tree constraints including the old feature can be updated to involve the new, and the old feature can be removed from the model.

### 5.3 updateall feature

The `updateall feature` command performs the same task with the `update feature` command, however, it can process multiple features unlike the `update feature` command, which can process exactly one feature. It has the following form:

`updateall feature` *Feature-Variable*
  `set` *Attribute-Updates*
  [*Where-Clause*] `;`

***Scenario***: Being satisfied by the sales figures for the service *"Dating Club"* after the extra cost reduction, marketing strategists come up with a maximum value of 5 USD a service or app can demand as an extra cost, in order to enlarge the market share. Therefore, the feature model must be revised and all features that do not conform to this new rule must be updated. ■

Multiple features can be updated with a single `updateall feature` command. This command will apply the changes to all of the features described. For instance, the following command will update the extra cost of all features that demand an extra cost higher than 5 USD, to the value 5 USD.

```
updateall feature F
  set extracost = numeric: 5
  where F.extracost > 5;
```

Similar to a single feature update command, this command can be used to update any of the attributes a group of features have, except for the `_decompID` and `_name` structural attributes, since, obviously two or more features cannot have the same name in a feature model. Any number of features can match the feature description to be updated, however, the command must not be ambiguous about the new value an attribute will receive, otherwise, it will have no effect.

***Scenario***: When the sales for the basic packages decrease marketing personnel launch an investigation and find out that the reason for the decrease is the services/applications demanding extra cost (e.g., parents who are not willing to pay extra amounts cancel their children's subscription to such packages). Thus, company executives decide to move all services/applications demanding extra amounts to *"Package 3"*. ■

The requirement that is presented in the above scenario can be fulfilled with the following command:

```
updateall feature F
  set _parent = "Package 3",
      _decomp = or to G
  where F.extracost > 0
   and (F._parent = "Package 1" or
        F._parent = "Package 2")
   and  G._parent = "Package 3"
   and  G.stype = "fun";
```

A multiple decomposition relations update command must be unambiguous about what the new parent will be and the specifics about the new decomposition relation, otherwise, the command will have no effect on the model. The features, whose decomposition relation with their parents will be updated, are described with a feature variable. This variable can resolve into a set of features, where performing the instructed task for some of the features from the resolution set will violate the feature model hierarchy (e.g., trying to move the root feature). If such a case arises, then the command will have a partial effect: instructed transformation will be applied to only the features that do not cause violations.

Multiple features update command does not add semantic power to Feather; the same effect can be obtained with a series of feature update commands. It has been added to the language with the purpose of obtaining increased convenience.

### 5.4 remove feature

Feature models can shrink by the removal of existing features. The `remove feature` command can be used to remove a feature from the model. It has the following form:

> `remove feature` *Feature-Descriptor*
> [*Where-Clause*] `;`

***Scenario (feature removal)***: Company X tech personnel discover that the component *"Video Chat"* does not provide high quality service and many user complaints had been received during the last couple of months. They decide to remove this component from the platform and introduce an enhanced version. Therefore, the feature model must not include the feature for this component anymore. ■

Modeler can achieve this goal with a `remove feature` command. For instance, the following command can be used to remove the feature *"Video Chat"* from the model:

```
remove feature "Video Chat";
```

This command type is designed to remove a single feature from the model, thus, it must not be ambiguous about which feature is to be removed (i.e., if a feature variable is used to describe the feature to be removed, then that variable must resolve into exactly one feature), otherwise, the command will have no effect. Similarly, a feature removal command must not instruct the removal of the root feature, which will destroy the entire model, otherwise, it will have no effect.

Removing a feature implies removing the entire subtree originating from the removed feature. For instance, if a command removes the feature *"Package 3"*, then the effects of the command will include removing all five features descending from it (*"Stock Wizard"*, *"Money Money Money"*, etc.) as well. Removing a feature also implies removing all cross-tree constraints involving the removed feature or one of its descendants. For instance, if the model includes the cross-tree constraint *"Video Chat" requires "High Speed Connection Protocol"* and a command removes the feature *"Video Chat"*, then the effects of the command will also include the removal of this constraint from the model.

### 5.5 removeall feature

The `removeall feature` command performs the same task with the `remove feature` command, however, it can process multiple features unlike the `remove feature` command, which can process exactly one feature. It has the following form:

> `removeall feature` *Feature-Variable*
> [*Where-Clause*] `;`

***Scenario***: When the overall income gained from the platform decreases, marketing personnel decide to reorganize the feature model. They decide to remove some of the components from the cheapest package, in order to force the users subscribing to packages that are more expensive. Thus, they decide to remove utility services from the *"Package 1"*. ■

The transformation requested by the marketing personnel can be performed using a `removeall feature` command.

```
removeall feature F
  where F._parent = "Package 1"
   and  F.stype = "utility";
```

This command will remove two features (*"My Way or Highway"* and *"Highway Jam"*) from the model. This command type does not add semantic power to Feather, the same effect can be obtained with a series of single feature removal commands. It has been added to the language with the purpose of obtaining increased convenience.

The features to be removed are represented with a feature variable in a multiple features removal command. This variable can resolve into a set of features that includes the root feature. If such a case arises, then the command will have a partial effect: all features except the root feature in the resolution set will be removed from the model.

### 5.6 add constraint

Transformations that can be applied to feature models are not limited to feature or decomposition relation manipulations. Adding cross-tree constraints can be required as well. Feather provides the `add constraint` command for this purpose. It has the following form:

`add constraint` *Constraint-Description*
[*Where-Clause*] `;`

The *Constraint-Description* has the following form:

*Feature-Descriptor-1*
*Constraint-Type Feature-Descriptor-2*

*Constraint-Type* can be `requires` or `excludes`. An `add constraint` command can add any number of cross-tree constraints to the model. If duplicate cross-tree constraints come into existence after the execution of an `add constraint` command, then the redundant copies are eliminated from the model.

***Scenario***: Tech personnel analyze the service requirements and discover that the *"Video Chat"* service requires a fast internet connection to provide high-definition image quality, therefore, ask the modeler to ensure this requirement is met. Then, when the financial services expand their coverage to include the worldwide markets, a similar requirement arises for them since they must make fast connections to get real-time data from several countries. ■

Following `add constraint` command can be used to perform the first part of the request in the above scenario:

```
add constraint "Video Chat" requires
     "High Speed Connection Protocol";
```

The second part requires multiple constraints to be added, which can be done by a command of the same type. For instance, the following command can answer the second part of the request:

```
add constraint F requires
      "High Speed Connection Protocol"
  where F._parent = "Package 3"
   and  F.stype = "utility";
```

### 5.7 update constraint

The `update constraint` command can be used to update exactly one cross-tree constraint in the model and has the following form:

`update constraint` *Constraint-Description*
`set` *Constraint-Updates*
[*Where-Clause*] `;`

The *Constraint-Updates* is a comma-separated list of constraint element updates that have the following form:

*Constraint-Element = value*

A *Constraint-Element* is a `leftfeature`, `constrainttype`, or `rightfeature`. The value that can be assigned to the `leftfeature` and `rightfeature` can be a *Feature-Descriptor*, and the value that can be assigned to the `constrainttype` can be a *Constraint-Type*.

Feature variables can be used to describe the features that figure in the existing constraint or the features that will figure in the updated constraint. A constraint update command cannot be ambiguous about the new features that will figure in the updated constraint. Since this command is specifically designed to update a single cross-tree constraint, it must not be ambiguous about which constraint is to be updated as well. If the description of the constraint to be updated matches two or more existing constraints, then the command will have no effect on the model.

***Scenario***: When the engineers working for the company develop a new protocol specifically designed for video transfer, the company adds a feature (i.e., *"Video Protocol"*) representing this component. Naturally, they ask the constraints to be modified to reflect this new situation. ■

An `update constraint` command can be used to modify an existing cross-tree constraint in order to adapt to the changes.

```
update constraint
     "Video Chat" requires
     "High Speed Connection Protocol"
  set rightfeature = "Video Protocol";
```

If duplicate cross-tree constraints come into existence after the execution of an `add constraint` command, then the redundant copy is eliminated from the model.

### 5.8 updateall constraint

The `updateall constraint` command can be used to update multiple cross-tree constraints with a single command and has the following form:

`updateall constraint` *Constraint-Description*
`set` *Constraint-Updates*
[*Where-Clause*] `;`

Similar to a constraint update command, a multiple constraints update command must not be ambiguous about the new features that will figure in the updated constraints. This command type does not add semantic power to Feather, the same effect can be obtained with a series of constraint update commands. It has been added to the language with the purpose of obtaining increased convenience.

Multiple copies of the same cross-tree constraint or multiple cross-tree constraints of the same effect (e.g., *"X" excludes "Y"* and *"Y" excludes "X"*) can come into existence as a result of executing a constraint(s) addition/update command. When such a situation occurs, redundant constraints are eliminated from the model so that the transformed model includes only a single instance of such constraints.

***Scenario***: When the company engineers develop a new component with better connection speed, a feature representing this component, *"Ultra Speed Protocol"*, is added to the model. Hence, cross-tree constraints must also be updated to ensure that better service quality is provided. ■

The following command can be used to perform the tasks requested in the above scenario:

```
updateall constraint
     F requires
     "High Speed Connection Protocol"
  set rightfeature = "Ultra Speed Protocol"
  where F._parent = "Package 3"
   and  F.stype = "utility";
```

### 5.9 remove constraint

A cross-tree constraint can be eliminated from a model with a `remove constraint` command. This command has the following form:

> `remove constraint` *Constraint-Description*
> [*Where-Clause*] `;`

Feature variables can be used to describe the constraint to be removed (e.g., *remove constraint F excludes "All Sideways" where …*). However, since this command is devised for removing a single constraint, the constraint description must not be ambiguous. If the description matches two or more existing cross-tree constraints, then the command will have no effect on the model.

***Scenario***: There is a lawsuit between the developers of the components represented by the features *"Highway Jam"* and *"All Sideways"*, which prevents both services to be used by the same user. This case is represented with an excludes cross-tree constraint in the model (i.e., *"Highway Jam" excludes "All Sideways"*). When the developers reach an agreement this constraint becomes void, so the company decides to eliminate it. ■

The following command can be used to remove the cross-tree constraint mentioned in the above scenario:

```
remove constraint
   "Highway Jam" excludes
   "All Sideways";
```

### 5.10 removeall constraint

When there is a need to remove multiple cross-tree constraints from a model, it can be performed with a `removeall constraint` command. This command has the following form:

> `removeall constraint` *Constraint-Description*
> [*Where-Clause*] `;`

The `removeall constraint` command performs the same task with the `remove constraint` command, however, it can process multiple cross-tree constraints. For instance, the following command will remove all of the cross-tree constraints from the model such that *"All Sideways"* figures in as an excluding feature:

```
removeall constraint
   F excludes "All Sideways";
```

This command type too, similar to other multiple entity processing commands, does not introduce additional semantic power. The same effect can be achieved via a series of constraint removal commands.

## 6. Semantics

### 6.1 Representations

This subsection presents the representations of the values that will be used while providing semantics for the commands. Two feature model related data types are used in the presentation: $T_{DR}$ = {*mandatory*, *optional*, *alternative*, *or*} and $T_{CTC}$ = {*requires*, *excludes*}.

A feature model is defined as a tuple *M* = (*F*, *D*, *C*), where

- *F* is the set of features in *M*
- *D* is the set of decomposition relations in *M*
- *C* is the set of cross-tree constraints in *M*

The root feature in *M* is represented with *root*(*M*). *name*(*f*) represents the name of the feature *f* and *names*(*F*) represents the set of names of the features in *F*. *subtree*(*M*, *f*) represents the set of all features {$f$, $f_1$, …, $f_n$}, which are included in the sub-tree that has *f* as its root in *M*. *parent*(*d*), *decType*(*d*), *child*(*d*), and *decIDNo*(*d*) represent the parent feature, decomposition relation type, the child feature, and the identification number assigned to this decomposition relation in the model, for the decomposition relation *d*, respectively. *leftF*(*c*), *ctcType*(*c*), and *rightF*(*c*) represent the left feature, cross-tree constraint type, and the right feature figuring in the cross-tree constraint *c*, respectively. *involvingCTCs*(*M*, *f*) represents the set of all cross-tree constraints in *M* that involves the feature *f* as its left or right feature.

*replacement*(*exp*, ($V_1$, …, $V_n$), ($f_1$, …, $f_n$)) represents the expression *exp'* obtained by replacing all occurrences of the feature variable $V_i$ by $f_i$ in the expression *exp*, for i = 1, …, n. *evaluation*(*M*, *exp*) represents the value obtained by evaluating the expression *exp* with respect to the feature model *M*. *resolution*(*M*, ($V_1$, …, $V_n$), *exp*) represents the set of all resolutions {$R_1$, …, $R_m$} for the feature variables ($V_1$, …, $V_n$) with respect to the feature model *M*, where $R_i$ = ($f_{i-1}$, …, $f_{i-n}$) and *evaluation*(*M*, *replacement*(*exp*, ($V_1$, …, $V_n$), $R_i$)) = *true*, for i = 1, …, m.

*describedF*(*M*, *FDesc*, *exp*) represents the set of all features described by *FDesc* in the feature model *M*, and is equal to: *(i)* {*f*}, if *FDesc* is a feature *f* such that *f* ∈ *F*, *(ii)* {$f_1$, …, $f_n$}, if *FDesc* is a feature variable and *resolution*(*M*, (*FDesc)*, *exp*) = {($f_1$), …, ($f_n$)}, and *(iii)* ∅, otherwise.

*describedC*(*M*, *CDesc*, *exp*) represents the set of all cross-tree constraints in the feature model *M* described by *CDesc* = $FDesc_1$ *CType* $FDesc_2$, where *CType* ∈ $T_{CTC}$, and is equal to: *(i)* {*c*}, if $FDesc_1 = f_1$, $f_1 \in F$, $FDesc_2 = f_2$, $f_2 \in F$, *leftF*(*c*) = $f_1$, *ctcType*(*c*) = *CType*, *rightF*(*c*) = $f_2$, and *c* ∈ *C*, *(ii)* {$c_1$, …, $c_m$}, if $FDesc_1$ is a feature variable *V*, $FDesc_2 = f$, $f \in F$, resolution(*M*, (*V*), *exp*) = {($f_1$), …, ($f_m$), …, ($f_n$)}, $c_i$ = ($f_i$, *CType*, *f*) ∈ *C* for i=1, …, m, $c_j$ = ($f_j$, *CType*, *f*) ∉ C for j = m+1, …, n, *(iii)* similar to the previous case, except for $FDesc_1$ is a feature and $FDesc_2$ is a feature variable, *(iv)* {$c_1$, …, $c_m$}, if $FDesc_1$ is a feature variable $V_1$, $FDesc_2$ is a feature variable $V_2$, resolution(*M*, ($V_1$, $V_2$), *exp*) = {($f_{1-1}$, $f_{1-2}$), …, ($f_{m-1}$, $f_{m-2}$), …, ($f_{n-1}$, $f_{n-2}$)}, $c_i$ = ($f_{i-1}$, *CType*, $f_{i-2}$) ∈ *C* for i=1, …, m, $c_j$ = ($f_{j-1}$, *CType*, $f_{j-2}$) ∉ C for j = m+1, …, n, *(v)* ∅, otherwise.

Cross-tree constraints that have the same effect are considered to be identical. For instance, if $c_1$ = ($f_1$, *excludes*, $f_2$) and $c_2$ = ($f_2$, *excludes*, $f_1$), then {$c_1$} ∪ {$c_2$} equals {$c_1$} (or {$c_2$}), and {$c_1$} – {$c_2$} equals ∅.

### 6.2 Command Semantics

A *feature addition* command requires the name of the feature to be added, description of the feature that will be the parent of the added feature, type of the decomposition relation between the new feature and its parent, description for a sibling feature, attribute declarations, and a where clause to guide the feature variable resolutions. Unneeded arguments (e.g., a sibling description when the new feature will not join in an existing group decomposition relation) can be null.

**addF**(*M*, *"name"*, *Par*, *Dec*, *Sib*, *AD*, *where*) = *M '*

It transforms the feature model *M* = (*F*, *D*, *C*) to *M '* = (*F'*, *D'*, *C'*), denoted by $M \xrightarrow{addF} M'$, by adding the new feature *f*, *F'* = *F* ∪ {*f*}, where *name*(*f*) = *"name"* and *"name"* ∉ *names*(*F*). *f* must include all the attribute declarations, which must not contain any ambiguities in the values to be assigned. The set of cross-tree constraints remains unchanged, *C'* = *C*. A new decomposition relation, *d*, is added to the set of decomposition relations, *D'* = *D* ∪ {*d*}, with the following characteristics.

The description for the parent feature that will figure in the decomposition relation must match exactly one feature in *F*, *parent*(*d*) = *p* such that *describedF*(*M*, *Par*, *where*) = {*p*}. The decomposition type, *decType*(*d*), is equal to: *(i) Dec*, if *Dec* ∈ $T_{DR}$, *(ii) decType*($d_1$), if it is described via a feature description (i.e., *Dec* is in the from *FDesc._decomp*), where *describedF*(*M*, *FDesc*, *where*) = {$f_1$, …, $f_m$}, *child*($d_1$) = $f_1$, …, *child*($d_m$) = $f_m$, and *decType*($d_1$) = … = *decType*($d_m$). Obviously, child feature in the relation is the added feature, *child*(*d*) = *f*. Relation identification number, *decIDNo*(*d*) is equal to: *(i)* 0, if this is a solitary relation (i.e., *mandatory* or *optional*), *(ii) decIDNo*($d_{sib1}$), if the new feature joins in the same group relation (i.e., *alternative* or *or*) with its sibling $s_1$ such that *describedF*(*M*, *Sib*, *where*) = {$s_1$, …, $s_n$}, *child*($d_{sib1}$) = $s_1$, *decType*($d_{sib1}$) = *Dec*, and all the described siblings join in the same relation (i.e., *child*($d_{sib2}$) = $s_2$, …, *child*($d_{sibn}$) = $s_n$, *decType*($d_{sib1}$) = … = *decType*($d_{sibn}$), and *decIDNo*($d_{sib1}$) = … = *decIDNo*($d_{sibn}$)), *(iii)* an unused number in the model, if the new feature joins in none of the existing group relations.

A *feature update* command requires a description for the feature to be updated, the new name of the feature, the update information for the desired attributes, a description for the feature that will be the new parent, the type of the new decomposition relation between it and its parent, a description for a sibling feature, and a where clause to guide feature variable resolutions. Unneeded arguments (e.g., a parent description, if the location of the feature remains unchanged) can be null.

**updateF**(*M*, *FDesc*, *Name*, *AU*, *Par*, *Dec*, *Sib*, *where*) = *M '*

It transforms the feature model *M* = (*F*, *D*, *C*) to *M '* = (*F'*, *D'*, *C'*), denoted by $M \xrightarrow{updateF} M'$, by updating the described feature in the specified manner. The command must be unambiguous about the feature to be updated, *describedF*(*M*, *FDesc*, *where*) = {*f*}. The set of features and the set of cross-tree constraints remain unchanged, *F'* = *F* and *C'* = *C*. If there is a name update (i.e., *Name* is not null and not equal to *name*(*f*) in *M*), where *Name* ∉ *names*(*F* – {*f*}), then *name*(*f*) = *Name* in *M '*. If there is an attribute update, then the attributes of *f* in *M '* must have the updated values. If there is only a name or attribute update (i.e., *Par* and *Dec* are null), then the set of decomposition relations remains unchanged as well. Otherwise, it is updated in the following manner.

Obviously, the root feature cannot be the subject of the command if a decomposition relation update is involved, *f* ≠ *root*(*M*). First the old decomposition relation is removed from the model, $D_{tmp}$ = *D* – {$d_{old}$}, where *child*($d_{old}$) = *f*. Then, a new decomposition relation is added, *D'* = $D_{tmp}$ ∪ {*d*}, in the manner described above for adding a feature. If the location of the feature is updated (i.e., *parent*(*d*) ≠ *parent*($d_{old}$)), then it must also be true that *parent*(*d*) ∉ *subtree*(*M*, *f*), since the described update cannot violate tree-structured hierarchy of the model.

A *feature removal* command requires a description for the feature to be removed. The *where* expression can be null if not needed.

**remF**(*M*, *FDesc*, *where*) = *M '*

It transforms the feature model *M* = (*F*, *D*, *C*) to *M '* = (*F'*, *D'*, *C'*), denoted by $M \xrightarrow{remF} M'$, by removing the described feature. Command must be unambiguous about the feature to be removed, *describedF*(*M*, *FDesc*, *where*) = {*f*}. Root feature cannot be removed from the model, $f \neq root(M)$. Effects of removing the feature are propagated through the model as described below.

All features in the subtree that has the removed feature as its root are removed from the model as well. Hence, $F' = F - \{f, f_1, \ldots, f_n\}$, where $subtree(M, f) = \{f, f_1, \ldots, f_n\}$. Similarly, all the decomposition relations involving a removed feature are deleted, $D' = D - \{d, d_1, \ldots, d_n\}$ such that $child(d) = f$, $child(d_1) = f_1, \ldots, child(d_n) = f_n$. Finally, all the cross-tree constraints involving a removed feature are deleted, $C' = C - (C_f \cup C_1 \cup \ldots \cup C_n)$ where $C_f = involvingCTCs(M, f)$, $C_1 = involvingCTCs(M, f_1), \ldots, C_n = involvingCTCs(M, f_n)$.

Semantics for the *multiple features update* and *multiple features removal* commands can be given by adopting the semantics for feature update and feature removal commands, respectively.

A *constraint addition* command requires the description of the cross-tree constraint(s) that will be added to the model. The *where* expression can be null if not needed.

**addC**(*M*, *CDesc*, *where*) = *M '*

It transforms the feature model $M = (F, D, C)$ to $M' = (F', D', C')$, denoted by $M \xrightarrow{addC} M'$, by adding the described cross-tree constraint(s). The set of features and the set of decomposition relations remain unchanged, $F' = F$ and $D' = D$. The set of cross-tree constraints grow with the added cross-tree constraint(s), $C' = C \cup \{c_1, \ldots, c_n\}$, where $describedC(M, CDesc, where) = \{c_1, \ldots, c_n\}$.

A *constraint update* command requires a description for the cross-tree constraint to be updated, descriptions of the features that will be the new left and right features, the new constraint type, and a where clause to guide feature variable resolutions. Unneeded arguments can be null.

**updateC**(*M*, *CDesc*, *NewL*, *NewR*, *NewT*, *where*) = *M '*

It transforms the feature model $M = (F, D, C)$ to $M' = (F', D', C')$, denoted by $M \xrightarrow{updateC} M'$, by updating the described cross-tree constraint in the specified manner. Command must be unambiguous about the cross-tree constraint to be updated, $describedC(M, CDesc, where) = \{c\}$. The set of features and the set of decomposition relations remain unchanged, $F' = F$ and $D' = D$. The described constraint is removed from the model and an updated constraint is added, $C' = (C - \{c\}) \cup \{c'\}$.

The updated constraint has the following properties. $leftF(c')$ equals to: *(i)* $leftF(c)$, if *NewL* is null, *(ii)* $f_1$, if $describedF(M, NewL, where) = \{f_1\}$. Similarly, $rightF(c')$ equals to: *(i)* $rightF(c)$, if *NewR* is null, *(ii)* $f_2$, if $describedF(M, NewR, where) = \{f_2\}$. Finally, $ctcType(c')$ equals to: *(i)* $ctcType(c)$, if *NewT* is null, *(ii)* *NewT*, if $NewT \in T_{CTC}$.

A *constraint removal* command requires a description for the cross-tree constraint to be removed. The *where* expression can be null if not needed.

**remC**(*M*, *CDesc*, *where*) = *M '*

It transforms the feature model $M = (F, D, C)$ to $M' = (F', D', C')$, denoted by $M \xrightarrow{remC} M'$, by removing the described cross-tree constraint. Command must be unambiguous about the cross-tree constraint to be removed, $describedC(M, CDesc, where) = \{c\}$. The set of features and the set of decomposition relations remain unchanged, $F' = F$ and $D' = D$. The set of cross-tee constraint shrinks, $C' = C - \{c\}$.

Semantics for the *multiple constraints update* and *multiple constraints removal* commands can be given by adopting the semantics for constraint update and constraint removal commands, respectively.

# 7. An Interpreter for Feather

The interpreter that will be presented in this section is responsible for validating and executing Feather scripts in order to provide automated support. It has been developed in Java and packed into an executable jar file that is licensed under the BSD license. The executable jar file and the source files can be found in [48].

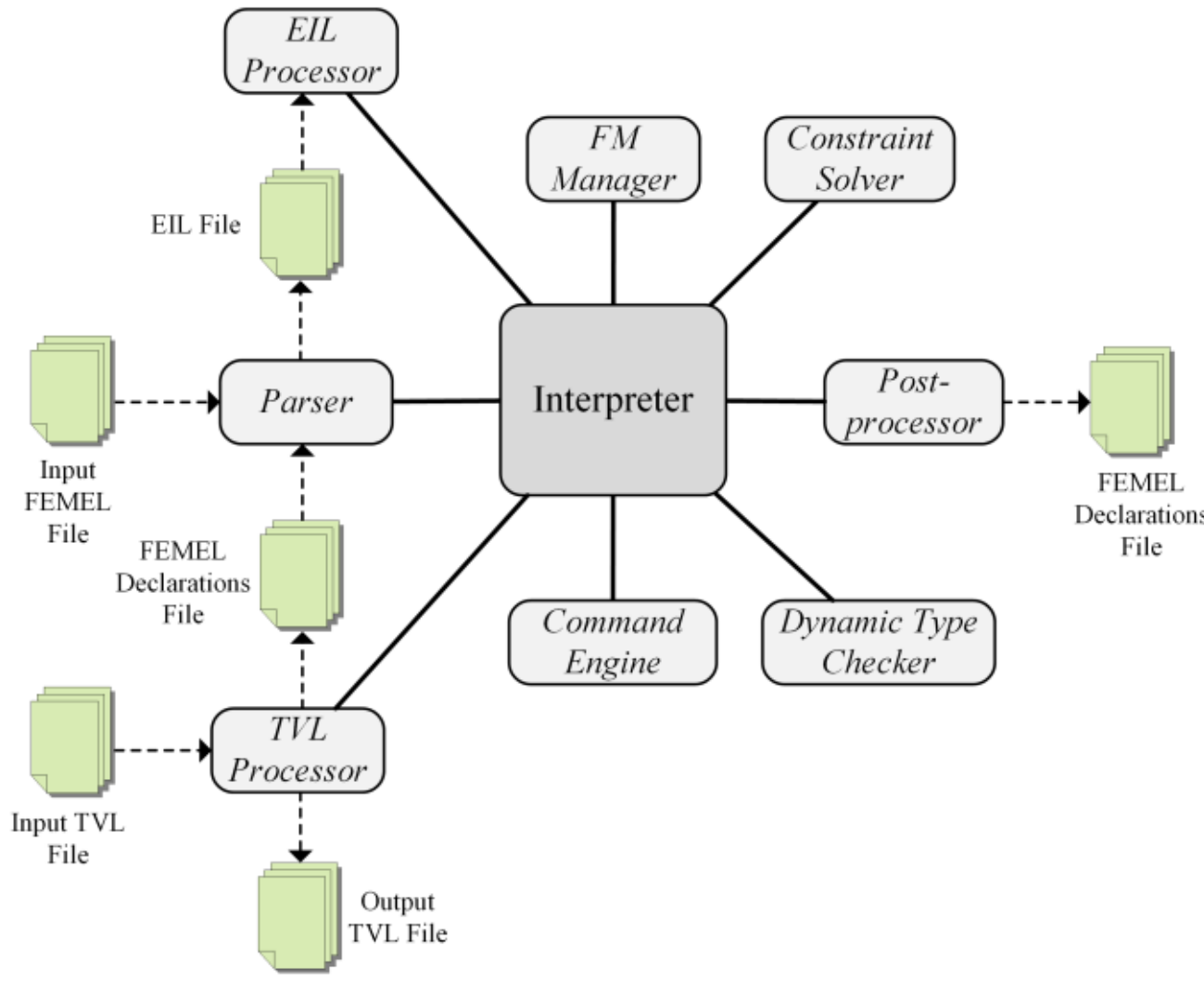

Fig. 4. Infrastructure of the interpreter

The interpreter incorporates an integrated parser to validate the input and several other components to perform the execution. Infrastructure of the interpreter is depicted in Fig. 4. Roles and orchestration of the components are discussed in the following subsections.

## 7.1 Input & Output

The interpreter takes its arguments and displays its messages in the command window. Arguments specify the running mode, the input script, and the name of the output file to be produced. Running mode indicates what to do when a warning or error is encountered while executing the commands. The interpreter has three modes: stop on the first warning/error, ignore warnings/stop on the first error, ignore all warnings/errors and continue running until all the commands are executed.

Input script can be fed to the interpreter as a single file that includes both the declarations and the commands, or two separate files, one for the declarations and one for the commands. If the input is a valid script, then the commands are executed on the input model. Finally, the *Post-processor* component generates the declarations code for the final model and saves in the output file.

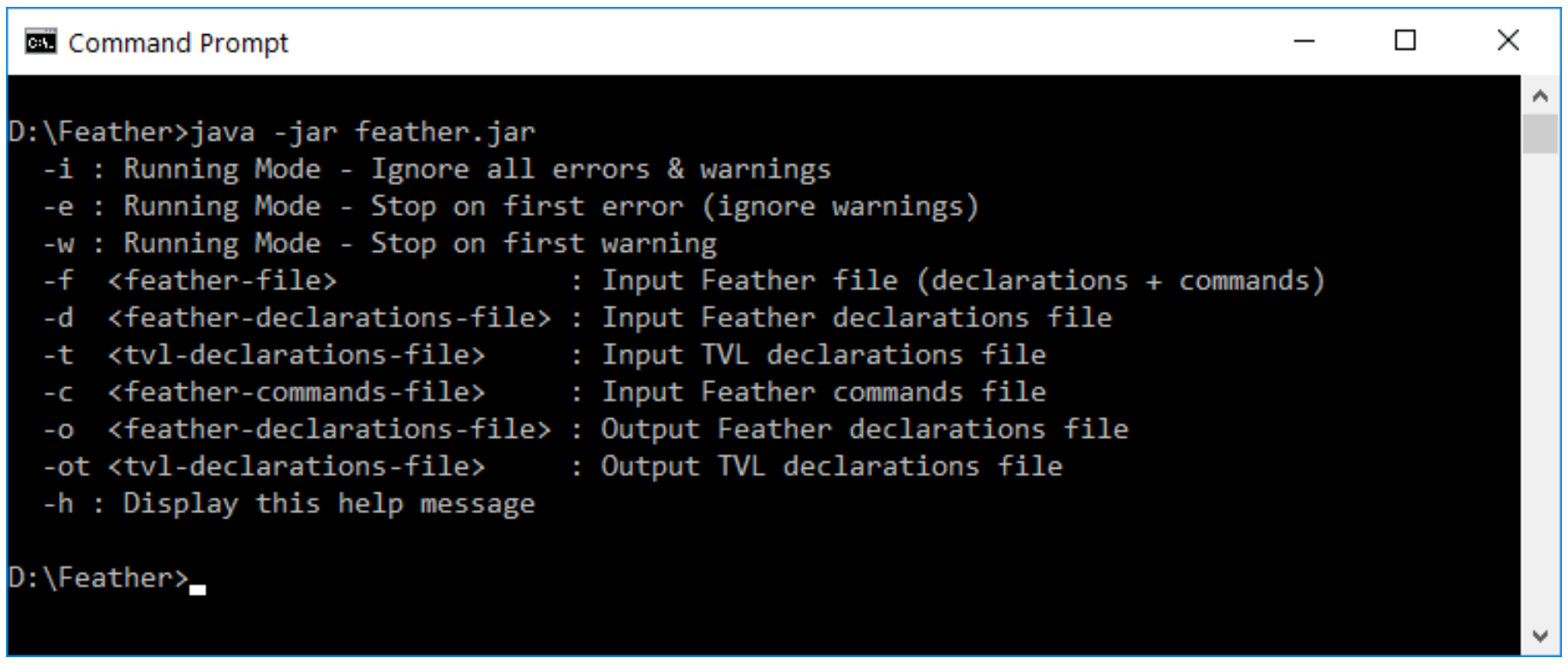

```
Command Prompt

D:\Feather>java -jar feather.jar
  -i : Running Mode - Ignore all errors & warnings
  -e : Running Mode - Stop on first error (ignore warnings)
  -w : Running Mode - Stop on first warning
  -f  <feather-file>              : Input Feather file (declarations + commands)
  -d  <feather-declarations-file> : Input Feather declarations file
  -t  <tvl-declarations-file>     : Input TVL declarations file
  -c  <feather-commands-file>     : Input Feather commands file
  -o  <feather-declarations-file> : Output Feather declarations file
  -ot <tvl-declarations-file>     : Output TVL declarations file
  -h : Display this help message

D:\Feather>_
```

Fig. 5.  Interpreter command line execution options

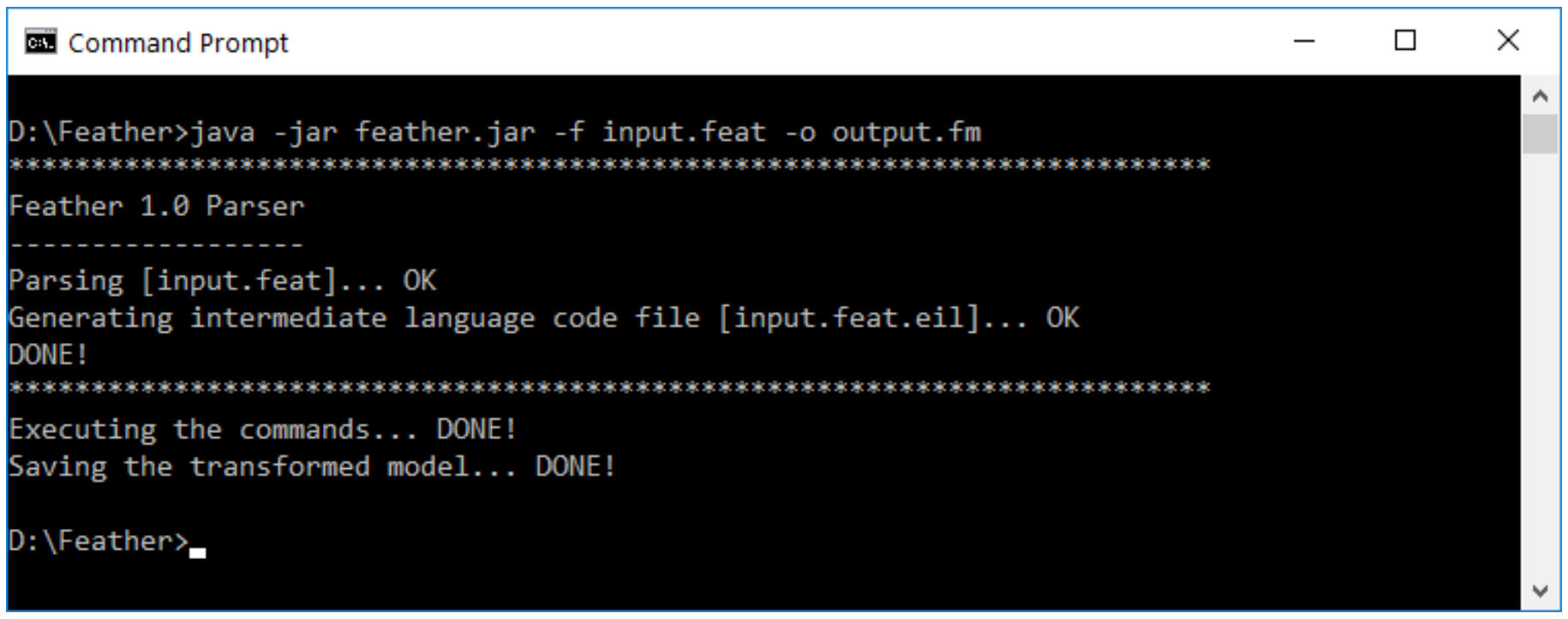

```
Command Prompt

D:\Feather>java -jar feather.jar -f input.feat -o output.fm
*************************************************************************
Feather 1.0 Parser
------------------
Parsing [input.feat]... OK
Generating intermediate language code file [input.feat.eil]... OK
DONE!
*************************************************************************
Executing the commands... DONE!
Saving the transformed model... DONE!

D:\Feather>_
```

Fig. 6.  A successful execution

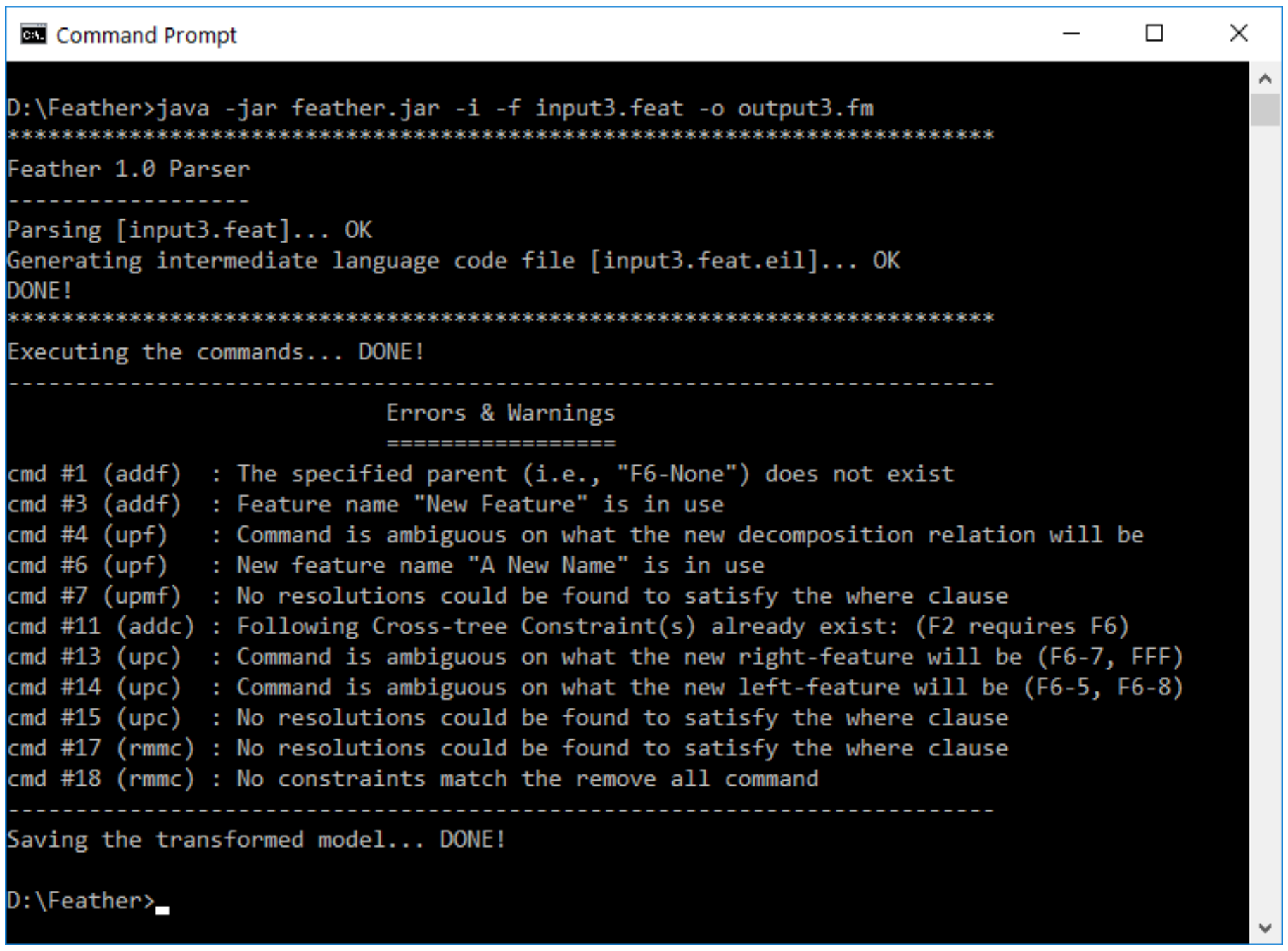

```
Command Prompt

D:\Feather>java -jar feather.jar -i -f input3.feat -o output3.fm
*************************************************************************
Feather 1.0 Parser
------------------
Parsing [input3.feat]... OK
Generating intermediate language code file [input3.feat.eil]... OK
DONE!
*************************************************************************
Executing the commands... DONE!
-------------------------------------------------------------------------
                            Errors & Warnings
                            =================
cmd #1 (addf)  : The specified parent (i.e., "F6-None") does not exist
cmd #3 (addf)  : Feature name "New Feature" is in use
cmd #4 (upf)   : Command is ambiguous on what the new decomposition relation will be
cmd #6 (upf)   : New feature name "A New Name" is in use
cmd #7 (upmf)  : No resolutions could be found to satisfy the where clause
cmd #11 (addc) : Following Cross-tree Constraint(s) already exist: (F2 requires F6)
cmd #13 (upc)  : Command is ambiguous on what the new right-feature will be (F6-7, FFF)
cmd #14 (upc)  : Command is ambiguous on what the new left-feature will be (F6-5, F6-8)
cmd #15 (upc)  : No resolutions could be found to satisfy the where clause
cmd #17 (rmmc) : No resolutions could be found to satisfy the where clause
cmd #18 (rmmc) : No constraints match the remove all command
-------------------------------------------------------------------------
Saving the transformed model... DONE!

D:\Feather>_
```

Fig. 7.  An example execution where a number of errors and warnings are generated

### 7.2 Parsing

The *Parser* component includes a lexer and a parser for the language. It parses and checks the input script against the context-free grammar of Feather to understand if the script is well formed or not. Skeleton of the component has been built on the Java code generated by the ANTLR parser generator tool [37], where some of the methods have been overridden in order to provide customized error handling.

Parser also performs several additional tasks such as; checking if there are repetitive declarations of features or attributes, if there are repetitive part updates in a feature or constraint update command (e.g., updating the left feature twice in a single constraint update command), if all the features included in the constraint declarations are declared before, etc. It uses an attribute grammar, where actions that utilize synthesized and inherited attributes are attached to the context-free grammar rules, to perform such checks.

Since Feather can be used by domain experts, who can be unfamiliar to programming, readability and simplicity have been considered as important factors during the language design. However, once an input script is validated against the grammar of the language, the remaining tasks are performed automatically, thus, readability and simplicity are no longer a concern. Therefore, an intermediate language, which is designed to found a basis of communication between the parser and the components responsible for execution, has been constructed.

This language is called the Evolution Intermediate Language (EIL) and used to express Feather declarations and commands for easier automatic handling. For instance, Feather allows arithmetic expressions to be coded in the infix notation, which is natural for humans, however, requires extra effort to determine the order of execution for the involved operations. On the other hand, the arithmetic expressions in EIL are coded in the postfix notation, which includes the order of execution information inherently.

Translating the validated input script from Feather to EIL is another responsibility of the parser. It performs this task using the aforementioned attribute grammar that also includes the necessary translation actions. The resulting translation is saved in a temporary file for the *EIL Processor* component.

The EIL processor is responsible from processing this translation file. It reads the translation file, sorts the declarations and the commands out, and stores them in appropriate internal data structures. Then, the component provides the input script elements to the interpreter engine.

Interpreter engine feeds the elements describing the input declarations to the *FM Manager* component, which is responsible for maintaining the feature model under transformation. FM manager uses these declarations to construct the model. Values for the `_decompID` structural attributes of the features, which are read-only values from the modeler's point of view, are assigned automatically during this phase. Once the model is constructed, interpreter becomes ready to execute the commands that will transform the feature model.

### 7.3 Constraint Solving

A major strength of Feather is allowing feature variables to be used while describing features. Modelers can utilize feature variables at programming time, where such variables must be resolved into actual features at runtime. The *Constraint Solver* component performs this resolution task during execution. This component is not a general constraint solver software; rather, it utilizes a third-party off-the-shelf tool, SICStus Prolog's jasper library [43], to achieve its goal.

Interpreter executes the commands one by one in the order they are provided. Since previous commands can change the structure of the feature model, and consequently the set of features a feature variable can resolve into, the task of finding resolutions for the variables in a command is performed right before executing the command. To achieve this goal, the constraint solver component performs a series of mappings to translate the problem in hand (i.e., finding resolutions for the feature variables) into a constraint program the third-party tool can process. The obtained constraint program consists of two main parts: the domain specifications and the conditions that must be satisfied.

Domain specifications establish the relations between the feature variables in the command and the variables in the constraint program. Each feature variable is represented by a constraint variable, where the constraint variable's domain consists of features the corresponding feature variable can resolve into. However, since *FeatureVariable.attribute* terms can figure in expressions in a command as well, additional constraint variables must also be used to represent a feature variable fully. Therefore, the component introduces two additional sets of constraint variables. The first set includes a constraint variable for each attribute of the feature variable to be resolved. The second set includes a constraint variable for each relevant attribute of the features the feature variable can resolve into. Finally, the constraint variables that belong to the same entity (i.e., the feature variable or the feature) are tied together. This strategy enables treating features and attributes, which reside on two different levels of abstraction, on the same footing.

For instance, assume that there is a feature variable *V*, two candidate features, *"F1"* and *"F2"*, this variable can resolve into, and the terms *V.a* and *V.b* figure in some expressions in the command. Then the mapping for this case becomes:

- $V_{cv} \in \{F1, F2\}$
- $(V_{vc} = F1 \Rightarrow V_a = F1_a \wedge V_b = F1_b)$
- $(V_{vc} = F2 \Rightarrow V_a = F2_a \wedge V_b = F2_b)$

The first item in the list specifies the domain of the constraint variable that represents the feature variable. The second and third items tie the additional constraint variables used to represent an entity together. All items are combined with logical conjunction in the constraint program.

A feature variable can represent any feature in the actual model; however, not every actual feature can substitute a feature variable. For instance, if there is an expression including the term *V.a*, then no actual feature that does not have an attribute named *a* can replace the feature variable *V*. Thus, the component does not insert such features in the domain of the constraint variable representing the feature variable *V* during the mapping process.

Having an attribute with the same identifier is necessary to be included in the domain of a constraint variable, however, not sufficient. The type of the attribute must also be compatible with the expression the feature variable figures in. For instance, assume that there is an expression *V.a < 5* in the command and a feature, *"F1"*, has an attribute named *a*. If the type of *a* is Boolean, obviously *V* must not resolve into *"F1"*, since otherwise the expression will not be well formed. Thus, a dynamic type check must also be performed while determining the candidate features that will constitute the domain of the constraint variable.

The *Dynamic Type Checker* component provides this service at runtime. It checks the set of candidate features and eliminates the ones that will create type inconsistency. This ensures that the feature variable will resolve into features whose attributes are compatible with the expressions the feature variable figures in. This elimination also increases the efficiency of the solution, since the third-party tool will have to deal with a constraint program that has smaller domains for its variables.

The second part of the obtained constraint program includes the conditions that will guide the resolution process. The conditions that must be satisfied by an actual feature that will replace a feature variable in a command are specified in the *Where-Clause* of the command. Thus, the component also maps this clause in order to guide the constraint solving process. Once the mapping process is complete, the component saves the translated constraint program in a temporary file, invokes the jasper tool to find the resolutions for the feature variables, and returns the resolution sets reported by the tool, if there exists any.

### 7.4 Command Execution

A command can involve feature variables, which must be resolved right before execution, as discussed in the previous subsection. A command can also involve *"Feature".attribute* terms, which must be checked for type consistency. However, this type check cannot be performed during parsing and has to be performed dynamically as discussed in the following paragraph.

Assume that a command includes an operation *"F".attr + 5* in an expression. This operation is considered well-formed if and only if: *(i)* there is a feature with the name *"F"*, *(ii) "F"* has an attribute *attr*, and *(iii)* the type of *attr* is numeric. Also assume that there is a feature, which satisfies all three conditions, in the declarations. Even this second assumption cannot guarantee that this operation is valid, since a previous command can remove the aforementioned feature and another command can add a new feature with the same name, which is vacated by the previous remove command, where the new feature has an attribute *attr* that has the type string. Therefore, type checking cannot be performed during parsing and must be performed dynamically when features are involved. The *Dynamic Type Checker* component performs these kinds of checks right before the execution, ensuring that the current structure of the feature model under transformation is taken into account for type checking at any time.

Once the feature variables (if there are any) are resolved into actual features and the command is validated for type consistency, the *Command Engine* component takes over the control to execute the command. It instructs the *FM Manager* component to update the feature model to reflect the transformation described by the command. Then, it reports the result of the execution attempt to the interpreter engine.

The *FM Manager* component performs various tasks automatically in order to ensure that the effects of the commands propagate as expected. For instance, when a feature is removed from the model, it automatically removes all the features descending from the removed feature and all the constraints including the removed feature(s). This component is also responsible from ensuring the feature model integrity. For instance, if a command tries to remove the root feature, which will destroy the entire model, component rejects to perform the task and reports an error. As another example, if a command tries to move a feature under one of its descendants, which will cause a cycle in the feature hierarchy, this component detects the problem and responds with an error.

### 7.5 A TVL Processor Extension

This extension is a component that adds the capability to process feature models that have already been expressed in TVL [14]. Since Feather includes all the declaration structures required to represent a feature model that will undergo transformation, this component is not a core part of the interpreter; it has been added in order to widen the potential user-base.

TVL is capable of expressing extended feature models, which makes it suitable for representing the feature models that will be transformed by the interpreter. However, TVL also includes some extra structures that are not allowed in Feather. For instance, it is possible to express complex cross-tree constraints in TVL, which are not allowed in Feather. Therefore, this component is capable of processing only a subset of the TVL. The context-free grammar of the allowed subset is provided in the appendices.

The first main task of this component is translating TVL expressions for the input feature model into Feather declarations. It facilitates a parser that has been developed using ANTLR for this translation process. This parser checks the validity of the input model representation according to TVL, and if the input is valid, performs the translation using the actions specified in the attribute grammar that has been constructed for this purpose. Then, the temporary output declarations file, which represents the input model in Feather, is fed to the interpreter.

The second task of this component is saving the resulting feature model that represents the transformed feature model in TVL. It generates the necessary TVL expressions from the internal data structure representing the transformed feature model and saves them in an output file.

## 8. Evaluation

This section presents two examples on how Feather can be used in two environments with different characteristics. The first environment is a highly dynamic environment, where a feature model is used to manage the variability in a mobile platform. Various applications and services are constantly installed/registered to and removed from the platform, hence, the structure of the feature model must continuously evolve. The second one is a relatively static environment that is about to undergo a drastic change. This one, too, uses a feature model to manage the variability among a large number of parts that can be used to build a computer configuration. In addition to these examples, results of a series of experiments conducted to assess the performance of the Feather interpreter are presented in the last subsection.

### 8.1 Case 1: A Dynamic Environment

Imagine a company developing a platform for mobile devices that will manage the applications and services installed/registered. Due to several reasons such as relatively limited resources available (e.g., processing power, memory, internet connection bandwidth) or inter-application/service constraints (e.g., two services needing the same resource to run) all installed products cannot be active simultaneously, hence, the platform needs to use a mechanism to model and manage variability in order to build valid running configurations. The company decides to use a feature model to achieve this goal.

Since there is a very rich repository of products for mobile devices, users can constantly install/register new applications/services and uninstall/unregister old ones. Thus, the internal structure of the feature model can constantly change and it must evolve continuously.

The first scenario in this environment discusses a situation from the point of view of a third-party application developer who wants her product to be installable to the platform. Assume that the product is a weather forecast application with the name *"Shining Sun"*. Developer would want to ensure that the new feature representing the product is added to the correct place in the feature model so that it will function as expected. However, the internal structure of the feature model may not be available to the developer due to several reasons (e.g., the company developing the platform can keep the structure classified due to commercial considerations). How would the developer overcome this problem?

The developer can write a Feather script to add the feature representing the product into the correct location by using only general guidelines published by the platform owner (e.g., *"The containers for the applications that require internet connection have an attribute named 'connected' with the value true"*). Developer uses a feature variable to describe where the new feature must be added to and leaves the task of positioning the feature correctly to the Feather interpreter. This script can look like the following:

```
add feature "Shining Sun"
  with attributes (
    _parent = F._name,
    _decomp = optional,
    type = inherited: F.type,
    price = 2.99)
  where F.varPoint = true
   and  F.category = "utility"
   and  F.subcategory = "weather forecast"
   and  F.connected = true
   and  F.minPrice <= 2.99
   and  F.maxPrice >= 2.99;
```

This script can work seamlessly regardless of the internal structure of the feature model. For instance, it would work fine in both of the feature models depicted in Fig. 8.

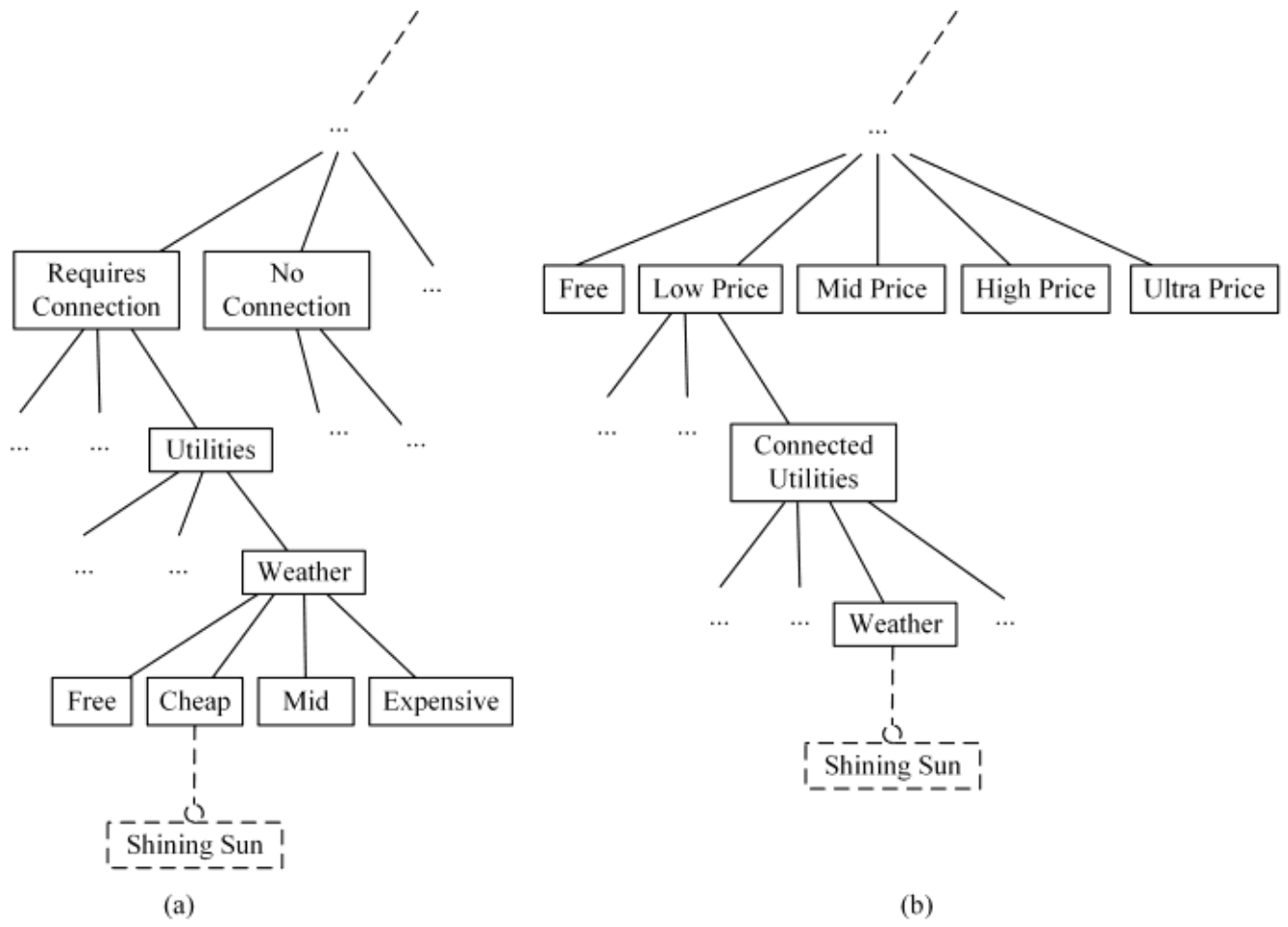

Fig. 8. Two possible feature models with different internal structures.

The second scenario in this environment discusses a situation from the point of view of the platform owner company. The company decides to change the structures of the feature models of all users, which can be different from user to user since different users' model structures can be different depending on the products installed/registered to. Assume that the company decides to cease support for a game genre (e.g., shooter games), hence, the variation point representing this genre must be deleted from the models. However, it can be the case that some of the users have already bought several games belonging to the genre, therefore, the variation point (and its descendants) cannot be removed directly. How can the company come up with a solution that will cover all these cases successfully?

The company can write a Feather script to move all games of the genre that are already installed, if there are any, and then delete the variation point. Company can use a feature variable to describe all such games and leave the task of figuring and relocating all of them dynamically for each individual user to the Feather interpreter. This script would look like the following:

```
updateall feature F
  set _parent = "Obsolete",
      _decomp = optional
  where F._parent = "Shooter Games";

delete feature "Shooter Games";
```

This script can work seamlessly regardless of the number of games already installed under this variation point. For instance, it can work successfully in both of the following feature models depicted in Fig. 9.

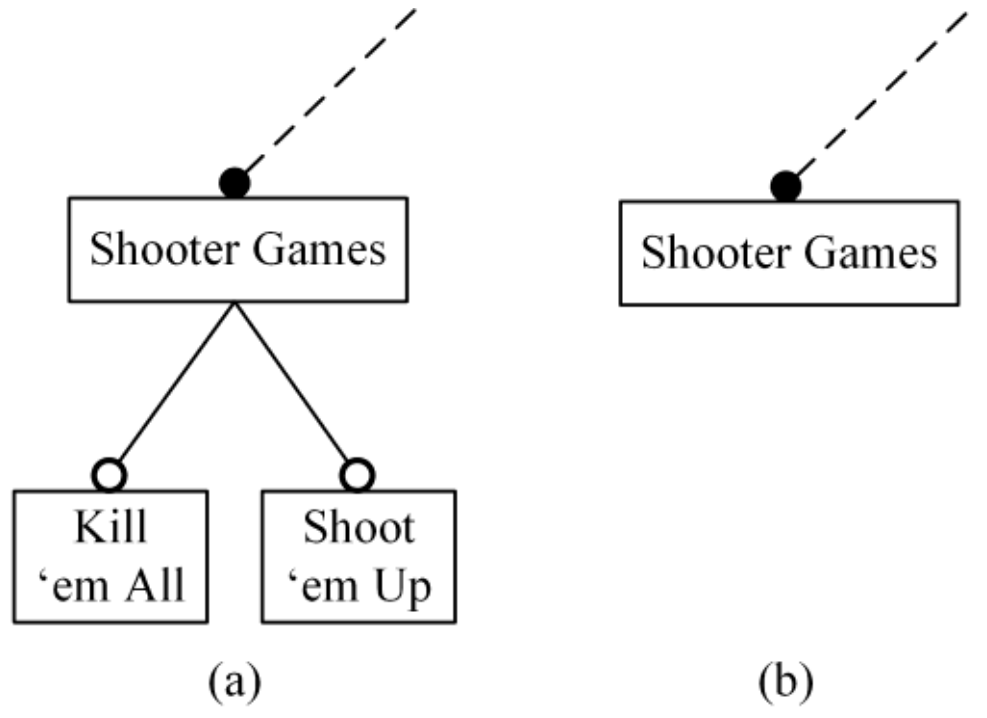

Fig. 9. The first user has two games of the genre, whereas the second has none.

### 8.2 Case 2: A Static Environment

Imagine a computer retailer company building custom computer configurations to meet various customers' needs by assembling appropriate parts from their stocks. The company uses a feature model to model the variability among the computer parts. Since the company wants to appeal to a wide customer base they keep many different computer parts in their stocks, and consequently, they have to manage a large feature model that consists of 1,227 features. Feature model diagram for the top-level features is depicted in Fig. 10. The complete feature model and the scripts can be found in [48].

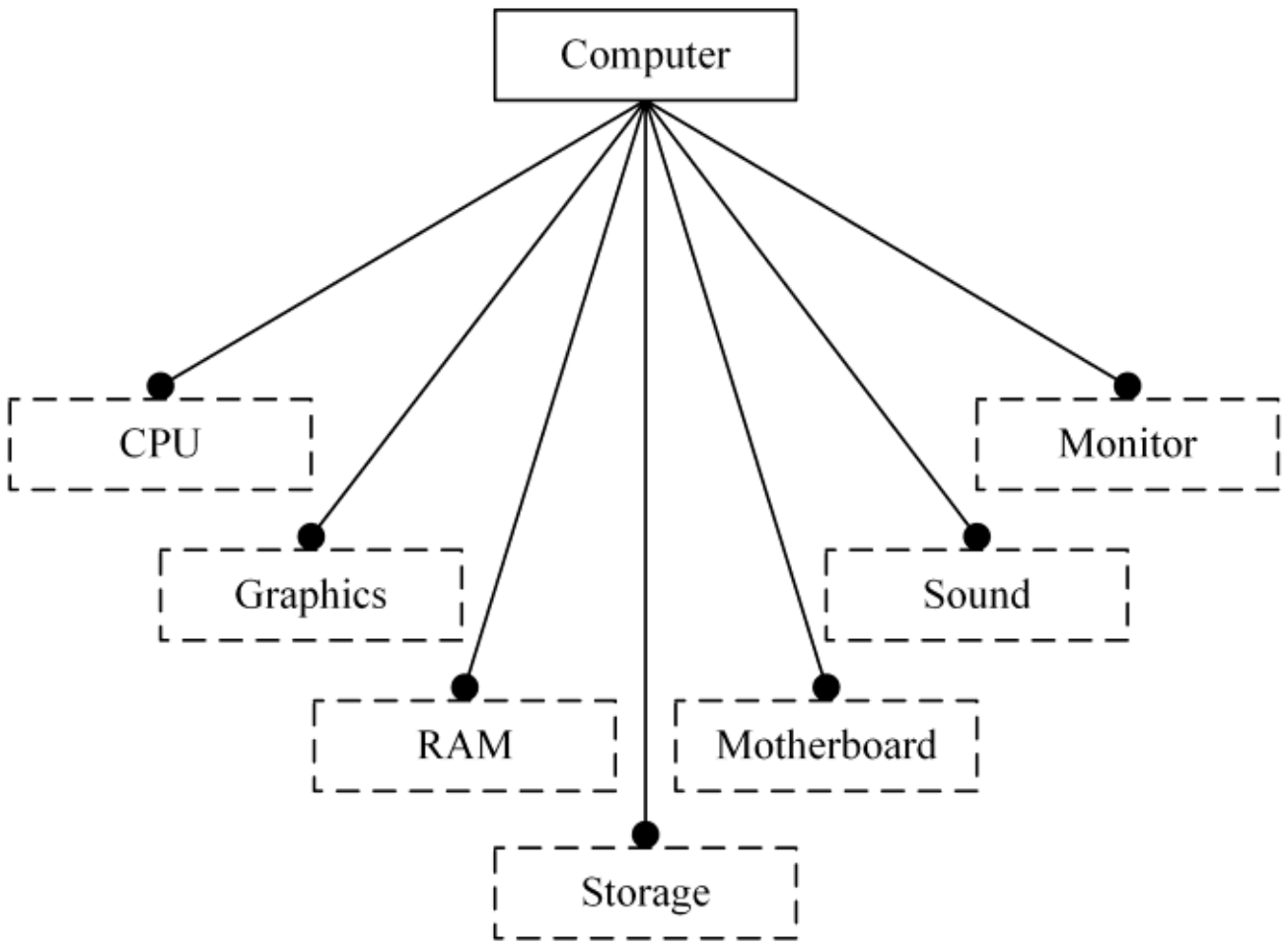

Fig. 10. Top level features of the computer parts feature model diagram.

Company decides to utilize a configuration assistant software to assist sales representatives while building a configuration for a customer. Therefore, the feature model must be restructured for the adaptation of the configuration assistant.

Branch A determines that cost is the single important factor in their customers' choices and decides to restructure the variability using decomposition relations. They decide to add five new features representing different price categories ("Pricing – Ultra", "Pricing – High", etc.) and move the features representing the computer parts under these categories. They write a Feather script to achieve this goal.

```
add feature "Configuration Assistant"
  with attributes (
    _parent = "Computer",
    _decomp = mandatory);

add feature "Pricing - Ultra"
  with attributes (
    _parent = "Configuration Assistant",
    _decomp = alternative,
    priceCategory = numeric : 5);

…

updateall feature F
  set _parent = "Pricing - Ultra"
  where F.priceCat = 5;

…

remove feature "CPU";

…
```

The script consists of 53 Feather commands to add new features for the configuration assistant, moving the features representing the computer parts, and removing obsolete variation points. It adds 6 new features, moves 1,154 features, and removes 72 features when executed. The whole process (i.e., parsing the input Feather file that includes the declarations and the commands, generating and processing the evolution intermediate language file, executing the commands, and saving the Feather declarations file for the resulting model) takes around 11 seconds on a notebook computer with an Intel i7-6700HQ processor and 16 GB RAM.

Branch B determines that there are multiple important factors that shape the customers' choices: cost and/or performance. Since there exist more than one factor, they decide that they cannot solve the problem using decomposition relations. Hence, they decide to use cross-tree constraints to guide the configuration assistant software. They, too, decide to add new features for different categories of price and for different categories of performance (e.g., "Performance – Ultra", "Performance – High", etc.), and then add appropriate cross-tree constraints. They write a Feather script to achieve this goal.

```
add feature "Configuration Assistant"
  with attributes (
    _parent = "Computer",
    _decomp = mandatory);

…

add feature "Performance - Ultra"
  with attributes (
    _parent = "CA - Performance",
    _decomp = alternative,
    perfMax = numeric : 200,
    perfMin = numeric : 80);

…

add constraint F excludes G
  where F.priceCategory <> G.priceCat;

add constraint F excludes G
  where G.rating > F.perfMax
   or   G.rating < F.perfMin;
```

The script consists of 13 Feather commands to add new features and 2 commands to add new cross-tree constraints. It adds 13 new features and 7,016 cross-tree constraints to the model when executed. The whole process takes around 4 seconds on the notebook computer aforementioned.

### 8.3 Performance

Since Feather can be used in highly dynamic environments where the underlying hardware may have limited capabilities, a series of experiments have been conducted to analyze the performance of the interpreter. The major contributing factor to the time required to perform a task is the usage of feature variables as it facilitates a third-party constraint solver in order to find the valid resolutions for the variables used. The test cases have been built targeting this issue.

For the tests, three arbitrary feature models of different sizes have been built: small, medium, and large. The small model has 30 features and 11 cross-tree constraints, the medium model contains 100 features and 33 cross-tree constraints, and the large model has 250 features and 88 cross-tree constraints, as given in Table 1.

Table 1
Test Model Sizes

| Model | # of Features | # of Cross-tree Constraints |
|---|---:|---:|
| Small | 30 | 11 |
| Medium | 100 | 33 |
| Large | 250 | 88 |

Four instances of each command type in the language have been designed for each feature model. Each instance performs the same task on the model (e.g., removing the same set of features from the model), however, the first instance includes no feature variables, the second only one, the third two variables, and the fourth includes three feature variables. Then, these instances of commands have been organized into command sets with respect to the number of feature variables used in the instances, which makes four sets for each feature model, and twelve command sets in total.

Each command set has been executed on the fresh copies of the models and the time elapsed to execute each command has been measured. This process is repeated ten times and the arithmetic means of the recorded values have been taken into account as the time required to perform the task for each instance. The tests have been conducted on a notebook computer with an Intel i7-6700HQ processor and 16 GB RAM. The results are presented in Table 2. All the time units in the table are milliseconds.

The results indicate that there are two major contributing factors to the time required to perform a task; the number of feature variables used in the command and the size of the feature model under transformation. For instance, adding a feature takes less than one millisecond regardless of the size of the model when no feature variables are used in the command. However, using three feature variables for the same task takes 3, 16, and 384 milliseconds for the small, medium, and large models, respectively.

## 9. Related Work

Capilla et *al.* [10] identify automatic changes in the structural variability as a challenge and address this issue, among many other issues such as automated reconfigurations, dynamic diagnosis of the variability model, etc., in the runtime variability (RunVar) model they present. RunVar supports adding, removing, and moving a variant via utilizing the super-types (i.e., a list of strings defined by the user to categorize and classify a system feature) of relevant features. Authors state that changing the variability defined by variation points is more difficult to automate and limit their approach to cover only variants. Feather allows any change in the structural variability that can involve variation points and/or variants. RunVar strategy is limited to compatibility checks between the super-types of related features, which are essentially string equality checks, for the determination of the location a new feature will be added to or an existing feature will be moved to. Feather allows specifying complex formulas involving any attribute of any feature for this purpose. RunVar supports structural changes involving a single feature at a time, whereas Feather enables changes involving multiple features to be performed with a single command. RunVar supports structural changes involving only the addition, removal, and relocation of features. Feather enables, in addition to features, addition, update, and removal of cross-tree constraints as well.

Table 2
Performance Results

| Command | **Small Model** *(30 features)* | | | | | | | | **Medium Model** *(100 features)* | | | | | | | | **Large Model** *(250 features)* | | | | | | | |
|---|---|---|---|---|---|---|---|---|---|---|---|---|---|---|---|---|---|---|---|---|---|---|---|---|
| | **# of Variables** | **Time** | **# of Variables** | **Time** | **# of Variables** | **Time** | **# of Variables** | **Time** | **# of Variables** | **Time** | **# of Variables** | **Time** | **# of Variables** | **Time** | **# of Variables** | **Time** | **# of Variables** | **Time** | **# of Variables** | **Time** | **# of Variables** | **Time** | **# of Variables** | **Time** |
| Adding a Feature | – | 0 | 1 | 0 | 2 | 1 | 3 | 3 | – | 0 | 1 | 0 | 2 | 6 | 3 | 16 | – | 0 | 1 | 3 | 2 | 6 | 3 | 384 |
| Updating a Feature | – | 0 | 1 | 2 | 2 | 2 | 3 | 6 | – | 0 | 1 | 3 | 2 | 6 | 3 | 6 | – | 0 | 1 | 3 | 2 | 12 | 3 | 318 |
| Updating Multiple Features | – | N/A | 1 | 0 | 2 | 1 | 3 | 3 | – | N/A | 1 | 0 | 2 | 12 | 3 | 15 | – | N/A | 1 | 3 | 2 | 68 | 3 | 680 |
| Removing a Feature | – | 0 | 1 | 3 | 2 | 3 | 3 | 6 | – | 0 | 1 | 0 | 2 | 3 | 3 | 12 | – | 0 | 1 | 0 | 2 | 34 | 3 | 218 |
| Removing Multiple Features | – | N/A | 1 | 0 | 2 | 0 | 3 | 3 | – | N/A | 1 | 0 | 2 | 3 | 3 | 6 | – | N/A | 1 | 0 | 2 | 6 | 3 | 128 |
| Adding Constraints | – | 0 | 1 | 0 | 2 | 3 | 3 | 15 | – | 0 | 1 | 3 | 2 | 6 | 3 | 15 | – | 0 | 1 | 0 | 2 | 15 | 3 | 65 |
| Updating a Constraint | – | 0 | 1 | 0 | 2 | 3 | 3 | 3 | – | 0 | 1 | 0 | 2 | 9 | 3 | 12 | – | 0 | 1 | 3 | 2 | 28 | 3 | 261 |
| Updating Multiple Constraints | – | N/A | 1 | 0 | 2 | 1 | 3 | 3 | – | N/A | 1 | 0 | 2 | 3 | 3 | 6 | – | N/A | 1 | 3 | 2 | 3 | 3 | 171 |
| Removing a Constraint | – | 0 | 1 | 1 | 2 | 3 | 3 | 3 | – | 0 | 1 | 0 | 2 | 6 | 3 | 15 | – | 0 | 1 | 3 | 2 | 6 | 3 | 250 |
| Removing Multiple Constraints | – | N/A | 1 | 0 | 2 | 1 | 3 | 3 | – | N/A | 1 | 0 | 2 | 3 | 3 | 6 | – | N/A | 1 | 0 | 2 | 6 | 3 | 412 |

Pleuss et *al.* [39] propose a model-driven approach for managing SPL evolution on feature level and describe a specific feature model called EvoFM to represent the evolution steps. They present a number of operators called EvoOperators to specify structural changes by the update of one or more properties of feature model elements. Authors report that they have used the Epsilon Transformation Language (ETL) [32] to implement the model transformations. Feather is powerful enough to describe and realize all model transformations involving any combination of EvoOperators, and more, in a simple and natural way for feature modelers. Since Feather is designed specifically for feature model transformation, a feature modeler can use it without having to face the difficulties of learning and using a new general model transformation language that are discussed through the end of this section.

Bürdek et *al.* [9] propose an approach for reasoning about product-line evolution using complex feature model differences. They compare the old and new versions of a feature model and specify the changes using edit operations on feature diagrams. Authors present a catalog of atomic edit operations to change the structure of a feature model and describe complex edit operations formed by the combination of atomic and/or other complex edit operations. They use the Henshin model transformation language [3], which is based on algebraic graph transformation, to implement the edit operations. Similar to the case discussed in the previous paragraph, Feather is powerful enough to implement all the edit operations described in a simple and intuitive way for feature modelers.

In the literature there are several works reported that elaborate on the effects of feature model evolution. Seidl et al. [42] present a conceptual basis for a system to maintain consistency of a feature model and feature mappings by co-evolving the mappings when the feature model evolves. Baresi and Quinton [6] propose an architecture to check the consistency of a feature model and to bind the running configuration to a new configuration space when the feature model evolves. Thüm et al. [45] present an algorithm for reasoning about feature model edits, which takes two versions of a feature model (one representing the model before the evolution and the other representing the model after the evolution) as input and computes the change classification. Xue et al. [47] propose a method to assist analysts in detecting changes in the structure of a feature model that has undergone evolution. Borba et al. [8] describe a number of refinement transformation templates on feature models such as adding a mandatory feature to the model and propose a theory of product line refinement for safe evolution of SPLs. Peng et al. [38] describe a number of variability evolution patterns such as *'a mandatory feature turning into an optional feature'* and propose a value-based variability evolution analysis method for evolution history analyses and future evolution prediction. Figueiredo et al. [20] elaborate on the capabilities of aspect-oriented programming mechanisms for providing modularity and stability to SPLs during feature model evolution. Lotufo et al. [35] analyze the evolution of the Linux kernel feature model and categorize the changes in the structure of the feature model.

These studies do not focus on describing and realizing the feature model evolution steps. Rather, they describe a list of allowed operations to change the feature model structure or evolution scenarios, and analyze the effects of applying those operations or scenarios. Feather is powerful enough to describe and realize all these operations and scenarios if the feature model that undergoes evolution conforms to the Feather metamodel.

Alves et al. [2] extend the traditional notion of refactoring to include feature models, where they define a refactoring as a change to the structure of a feature model in order to improve certain aspects of the SPL. They also present a refactoring catalogue that includes changes such as *'collapsing optional and or'*. Holdschick [23] reports challenges in the evolution of SPLs in the automotive domain. Author also presents a number of real life examples from the automotive domain that require a change in the structure of a feature model such as *'adding new alternative features to the feature model when a new cruise control operating concept is introduced'*. It is possible to describe and realize all the refactoring definitions and changes presented in these studies in Feather.

Model transformation is a key activity for model driven approaches in software engineering. It has gained considerable attention especially since the standardization efforts by the Object Management Group (OMG). The reader can refer to [26] for a detailed survey on model transformation languages and tools, here some of these languages will be discussed.

Languages can be designed to process models that conform to various metamodels. Most popular metamodeling approach is the Meta-Object Facility (MOF) by OMG and several languages (e.g., ATL [25], ETL, JTL [13], Tefkat [34], UML-RSDS [33], VIATRA [17]) support models that conform to MOF or a metamodeling approach that is based on MOF. There are also some other metamodeling languages used such as MOLA MOF by MOLA [27], VPM by VIATRA, and Genmodel by Henshin. Models to go under evolution with Feather must conform to basic feature model notation extended with feature attributes.

There are different styles adopted by various languages. Some languages (e.g., Tefkat, UML-RSDS, JTL) adopt a declarative approach, where the focus is on which elements to transform without directly specifying how this transformation must be performed. Languages that adopt an imperative style (e.g., SiTra [1], Kermeta [24]) focus on when and how the transformation must be executed. Several languages (e.g., GReAT [5]) adopt a graph-based approach based on algebraic graph transformation. Some languages employ a hybrid style to combine the strengths of different approaches, such as ATL (declarative and imperative), MOLA (imperative and graph-based), and VIATRA (declarative, imperative, and graph-based). Feather adopts a declarative approach.

Some languages (e.g., Henshin, Kermeta) mainly support in-place transformations, whereas some other (e.g., Tefkat, SiTra) support out-place transformations. Some languages (e.g., MOLA, ATL, VIATRA) can perform both types. Feather's main design intent is performing in-place transformations. Rule scheduling can also vary from language to language. Some languages (e.g., Tefkat) do not let users specify scheduling and leaves this task to the tool providing the automated support. Some languages (e.g., Henshin, SiTra) provide constructs that enable users to specify various scheduling strategies such as sequential, non-deterministic, prioritized, and recursion-oriented. Feather lets the users specify the transformation steps and the rules are executed in the sequence provided.

Akehurst *et al.* [1] point to some serious concerns regarding the human factor in model transformation that can be summarized as follows. There are dozens of model transformation frameworks reported in the literature and each of them requires the potential user to learn a different language, which possesses its own specific language details and peculiarities even if they are based on the same standard. Model transformation is intrinsically difficult and to write even a simple transformation in these frameworks can require a lot of learning time. The steep learning curve can be an inhibitive factor in the usage of many frameworks. The only difficulty in the model transformation does not reside in the implementation of a transformation. Specification and definition of a model transformation is a complex task that requires significant domain knowledge and understanding of the model(s) involved. Specifying and defining a transformation requires a different set of skills to those required for implementing a transformation.

Feather is designed to propose a solution to this problem in the transformation of feature models and its main contribution is to the field of software product lines, especially the dynamic ones. Feather's constructs are simple and natural to feature modeling. For instance, several languages employ complex pattern languages to describe the source or target elements. Feather includes only two description patterns (i.e., for features and cross-tree constraints) to describe the source elements in the very same form they are used in feature modeling, simple yet powerful enough to describe everything needed for the transformation of a feature model. It adopts a simple declarative style inspired from SQL, a language that has been used successfully for decades by people from a wide range of disciplines including non-programmers. Since domain experts working in the construction and maintenance of product lines can exhibit a similar diverse profile, the adopted style can be promising for Feather as well. Declarative style also has some other advantages regarding the human factor. For instance, declarative style is closer to the way modelers intuitively perceive a transformation and it hides complex transformation details behind a simple syntax [25]. Declarative transformation languages are generally considered to be more useful when the source and target models have similar structure and organization [32]. Design of Feather also intends to benefit from these conditions.

In addition to these, the Feather interpreter provides built-in mechanisms to ensure feature model integrity during the transformation. For instance, if a command attempts to move a feature under one of its descendants, which will cause a cycle and a disconnected sub-graph, the interpreter automatically detects the violation and protects the model structure by not applying the transformation. Thus, modelers do not have to take additional precautions to ensure model integrity.

## 10. Conclusion

This article presents Feather, a feature model transformation language. Evolution of a feature model can be required as discussed by presenting several realistic examples throughout this article. Feather is proposed to contribute to the approaches addressing this challenge.

Feather is a novel model transformation language. However, its main contribution is not within the model transformation realm, as it is not intended to be an alternative to the existing languages for general model transformation tasks. Its main contribution resides in traditional and dynamic software product line evolution.

Performing the described transformations manually can require too much effort and be error-prone, which would make the procedure infeasible, hence, automated support is desirable. The presented Feather interpreter provides automated support to realize the described transformations effectively. It benefits from or incorporates mature third party tools such as ANTLR and SICStus Prolog, which have proven to be effective and efficient in their respective domains, to increase efficiency. Results from the performance analyses and examples indicate that the achieved level of efficiency makes the interpreter eligible for use in classic and dynamic systems.

Automated support requires precise definitions and well-formedness rules for the models that will be transformed and the commands that will describe the transformations. The mature feature model architecture provides this information for the models. For the commands, syntactic well-formedness rules are given using the context-free grammar of the language, and semantics are provided via definitions grounded in the properties of feature models.

Feature variables, which can be used to describe features via their properties, enhance the expressive power and flexibility of Feather significantly. Modelers can facilitate feature variables to describe features and leave the task of resolving them into actual features to the interpreter. This strategy shifts the workload from the modeler to the interpreter while increasing flexibility. Feature variables also enable operating in unknown territory (i.e., when the exact structure of the feature model to evolve is not available) that would widen applicability.

Model transformation is an intrinsically difficult task and learning a transformation language can require considerable time and effort, thus, it is desirable to provide an easy to learn and use language that includes familiar and simple concepts to feature modelers. These factors have shaped the design decisions while constructing Feather. For instance, Feather has a declarative style that is inspired from SQL and includes constructs that are familiar and intuitive to feature modelers. SQL is a very popular language that has been used by many people, who can be non-programmers, from a wide variety of disciplines. Feather user-base can exhibit a similar profile, since it can also be used by domain experts that can be from a wide variety of disciplines. It is believed that the similarity to the SQL style, which has proven to be useful for over decades, can be helpful to reach the targeted user base.

Several enrichments can be made to enhance Feather. Allowed attribute types, which is limited to integer, real, Boolean, and string currently, can be enriched with data types such as collections (e.g., sets, ordered lists, mappings) or even user-defined types. The allowed feature model structure can be extended to cover other constructs such as cardinality-based models. Language can be extended with SQL-like aggregates (e.g., *min*, *max*, *sum*) to increase the expressive power. Analysis capabilities and decision-making mechanisms that will enable analyzing the effects of the commands before applying them to the model and taking necessary actions depending on the results of the analyses (e.g., skipping a command if it will transform the model into a void model that represents no valid configurations) can be added to the interpreter.

## Acknowledgments

This work was supported by the Scientific and Technological Research Council of Turkey (TÜBİTAK) under Grant 215E188

## References

[1] D. H. Akehurst, B. Bordbar, M. J. Evans, W. G. J. Howells, and K. D. McDonald-Maier, "SiTra: simple transformations in Java", in *Proc. MoDELS 2006*, Genova, Italy, 2006, pp. 351–364.

[2] V. Alves, R. Gheyi, T. Massoni, U. Kulesza, P. Borba, and C. Lucena, "Refactoring product lines", in *Proc. GPCE '06*, Portland, OR, USA, 2006, pp. 201–210.

[3] T. Arendt, E. Biermann, S. Jurack, C. Krause, and G. Taentzer, "Henshin: advanced concepts and tools for in-place EMF model transformations", in *Proc. MoDELS 2010*, Oslo, Norway, 2010, pp. 121–135.

[4] T. Asikainen, T. Mannisto, and T. Soininen, "Kumbang: a domain ontology for modelling variability in software product families", *Adv Eng Inform*, vol. 1, no. 1, pp. 23–40, Jan. 2007, 10.1016/j.aei.2006.11.007.

[5] D. Balasubramanian, A. Narayanan, C. van Buskirk, and G. Karsai, "The graph rewriting and transformation language: GReAT", in *Proc. GraBaTs 2006*, Natal, Brazil, 2006.

[6] L. Baresi and C. Quinton, "Dynamically evolving the structural variability of dynamic software product lines", in *Proc. SEAMS '15*, Florence, Italy, 2015, pp. 57–63.

[7] T. Berger *et al.*, "A survey of variability modeling in industrial practice", in *Proc. VaMoS'13*, Pisa, Italy, 2013.

[8] P. Borba, L. Teixeira, and R. Gheyi, "A theory of software product line refinement", *Theor Comput Sci*, vol. 455, pp. 2–30, Oct. 2012, 10.1016/j.tcs.2012.01.031.

[9] J. Bürdek, T. Kehrer, M. Lochau, D. Reuling, U. Kelter, and A. Schürr, "Reasoning about product-line evolution using complex feature model differences", *Autom Softw Eng*, vol. 23, no. 4, pp. 687–733, Dec. 2016, 10.1007/s10515-015-0185-3.

[10] R. Capilla, J. Bosch, P. Trinidad, A, Ruiz-Cortés, and M. Hinchey, "An overview of dynamic software product line architectures and techniques: observations from research and industry", *J Syst Softw*, vol. 91, pp. 3–23, May 2014, 10.1016/j.jss.2013.12.038.

[11] L. Chen and M. A. Babar, "A systematic review of evaluation of variability management approaches in software product lines", *Inform Software Tech*, vol. 53, no. 4, pp. 344–362, Apr. 2011, 10.1016/j.infsof.2010.12.006.

[12] L. Chen, M. A. Babar, and N. Ali, "Variability management in software product lines: A systematic review", in *Proc. SPLC'09*, San Francisco, CA, USA, 2009, pp. 81–90.

[13] A. Cicchetti, D. Di Ruscio, R. Eramo, and A. Pierantonio, "JTL: a bidirectional and change propagating transformation language", in *Proc. SLE 2010*, Eindhoven, The Netherlands, 2010, pp. 183–202.

[14] A. Classen, Q. Boucher, and P. Heymans, "A text-based approach to feature modelling: Syntax and semantics of TVL", *Sci Comput Program*, vol. 76, no. 12, pp. 1130–1143, Dec. 2011, 10.1016/j.scico.2010.10.005.

[15] P. Clements and L. M. Northrop, "Salion, Inc.: A software product line case study", Software Eng. Inst., Carnegie Mellon Univ., Pittsburgh, Pennsylvania, USA, Tech. Rep. CMU/SEI-2002-TR-038, Nov. 2002.

[16] P. Clements and L. Northrop, *Software product lines: practices and patterns*, Boston, MA, USA: Addison-Wesley, 2001.

[17] G. Csertan, G. Huszerl, I. Majzik, Z. Pap, A. Pataricza, and D. Varro, "VIATRA - visual automated transformations for formal verification and validation of UML models", in *Proc. ASE 2002*, Edinburgh, Scotland, UK, 2002, pp. 267–270.

[18] K. Czarnecki, T. Bednasch, P. Unger, and U. Eisenecker, "Generative programming for embedded software: an industrial experience report", in *Proc. GPCE 2002*, Pittsburgh, PA, USA, 2002, pp. 156–172.

[19] K. Czarnecki, S. Helsen, and U. Eisenecker, "Formalizing cardinality-based feature models and their specialization", *Softw Process Improve Pract*, vol. 10, no. 1, pp. 7–29, Mar. 2005, 10.1002/spip.213.

[20] E. Figueiredo et *al.*, "Evolving software product lines with aspects", in *Proc. ICSE '08*, Leipzig, Germany, 2008, pp. 261–270.

[21] S. Hallsteinsen, M. Hinchey, S. Y. Park, and K. Schmid, "Dynamic software product lines", *Computer*, vol. 41, no. 4, pp. 93–95, Apr. 2008, 10.1109/MC.2008.123.

[22] M. Hinchey, S. Park, and L. Schmid, "Building dynamic software product lines", *Computer*, vol. 45, no. 10, pp. 22–26, Oct. 2012, 10.1109/MC.2012.332.

[23] H. Holdschick, "Challenges in the evolution of model-based software product lines in the automotive domain", in *Proc. FOSD '12*, Dresden, Germany, 2012, pp. 70–73.

[24] J. M. Jézéquel, O. Barais, and F. Fleurey, "Model driven language engineering with Kermeta", in *Proc. GTTSE 2009*, Braga, Portugal, 2009, pp. 201–221.

[25] F. Jouault, F. Allilaire, J. Bézivin, and I. Kurtev, "ATL: A model transformation tool", *Sci Comput Program*, vol. 72, no. 1–2, pp. 31–39, Jun. 2008, 10.1016/j.scico.2007.08.002.

[26] N. Kahani, M. Bagherzadeh, J. R. Kordy, J. Dingel, and D. Varro, "Survey and classification of model transformation tools", *Softw Syst Model*, to be published, 10.1007/s10270-018-0665-6.

[27] A. Kalnins, J. Barzdins, and E. Celms, "Model transformation language MOLA", in *Proc. MDAFA 2003/2004*, Linköping, Sweden, 2004, pp. 62–76.

[28] K. Kang, S. Cohen, J. Hess, W. Novak, and S. Peterson, "Feature-oriented domain analyses (FODA) feasibility study", Software Eng. Inst., Carnegie Mellon Univ., Pittsburgh, Pennsylvania, USA, Tech. Rep. CMU/SEI-90-TR-21, Nov. 1990.

[29] K. Kang, S. Kim, J. Lee, and K. Kim, "FORM: A feature-oriented reuse method with domain-specific reference architectures", *Ann Soft Eng*, vol. 5, pp. 143–168, Jan. 1998, 10.1023/A:101898062.

[30] A. S. Karataş, H. Oğuztüzün, and Ali Doğru, “From extended feature models to constraint logic programming”, *Sci Comput Program*, vol. 78, no. 12, pp. 2295–2312, Dec. 2013, 10.1016/j.scico.2012.06.004.

[31] A. Kleppe, J. Warmer, and W. Bast, *MDA explained, the model-driven architecture: practice and promise*, Boston, MA, USA: Addison-Wesley, 2003.

[32] D. S. Kolovos, R. F. Paige, and F. A. C. Polack, “The Epsilon transformation language”, in *Proc. ICMT 2008*, Zürich, Switzerland, 2008, pp. 46–60.

[33] K. Lano and S. Kolahdouz-Rahimi, “Specification and verification of model transformations using UML-RSDS”, in *Proc. IFM 2010*, Nancy, France, 2010, pp. 199–214.

[34] M. Lawley and J. Steel, “Practical declarative model transformation with Tefkat”, in *Proc. MoDELS 2005*, Montego Bay, Jamaica, 2005, pp. 139–150.

[35] R. Lotufo, S. She, T. Berger, K. Czarnecki, and A. Wąsowski, “Evolution of the Linux kernel variability model”, in *Proc. SPLC 2010*, Jeju Island, South Korea, 2010, pp. 136–150.

[36] T. Mens and P. Van Gorp, “A taxonomy of model transformation”, *Electron Notes Theor Comput Sci*, vol. 152 pp. 125–142, Mar. 2006, 10.1016/j.entcs.2005.10.021.

[37] T. J. Parr and R. W. Quong, “ANTLR: A predicated-LL(k) parser generator”, *Software Pract Exper*, vol. 25, no. 7, pp. 789–810, Jul. 1995, 10.1002/spe.4380250705.

[38] X. Peng, Y. Yu, and W. Zhao, “Analyzing evolution of variability in a software product line: From contexts and requirements to features”, *Inform Software Tech*, vol. 53, no. 7, pp. 707–721, Jul. 2011, 10.1016/j.infsof.2011.01.001.

[39] A. Pleuss, G. Botterweck, D. Dhungana, A. Polzer, and S. Kowalewski, “Model-driven support for product line evolution on feature level”, *J Syst Softw*, vol. 85, no. 10, pp. 2261–2274, Oct. 2012, 10.1016/j.jss.2011.08.008.

[40] K. Pohl, G. Böckle, and F. Linden, *Software product line engineering: foundations, principles, and techniques*, Berlin-Heidelberg, Germany: Springer-Verlag, 2005.

[41] D. C. Sharp, “Component based product line development of avionics software”, in *Proc. SPLC-1*, Denver, CO, USA, 2000, pp. 353–369.

[42] C. Seidl, F. Heidenreich, and U. Aßmann, “Co-evolution of models and feature mapping in software product lines”, in *Proc. SPLC '12*, Salvador, Brazil, 2012, pp. 76–85.

[43] SICStus prolog, https://sicstus.sics.se/, *accessed March 2019*.

[44] M. Simos *et al.*, “Software technology for adaptable reliable systems (STARS) organization domain modeling (ODM) guidebook version 2.0”, Lockheed Martin Tac. Def. Sys., Manassas, VA, USA, Informal Tech. Rep. STARS-VC-A025/001/00, Jun. 1996.

[45] T. Thüm, D. Batory, and C. Kastner, “Reasoning about edits to feature models”, in *Proc. ICSE '09*, Vancouver, Canada, 2009, pp. 254–264.

[46] D. L. Webber and H. Gomaa, “Modeling variability in software product lines with the variation point model”, *Sci Comput Program*, vol. 53, no. 3, pp. 305–331, Dec. 2004, 10.1016/j.scico.2003.04.004.

[47] Y. Xue, Z. Xing, and S. Jarzabek, “Understanding feature evolution in a family of product variants”, in *Proc. WCRE 2010*, Beverly, MA, USA, 2010, pp. 109–118.

[48] Supplementary material, https://github.com/askaratas/Feather, *accessed March 2019*.

## Appendix A: Context-free Grammar of Feather

*S* ::= *Declarations Commands*

*Declarations* ::= *FeatureDeclarations* *CrossTreeConstraintDeclarations*

*FeatureDeclarations* ::= *RootFeature* *OtherFeatureDeclarations*
*RootFeature* ::= `root` *FeatureName* *AttributeDeclarations* ;
*OtherFeatureDeclarations* ::= ϵ
| *FeatureDec* *OtherFeatureDeclarations*
FeatureDec ::= `feature` *FeatureName* *ParentName* *DecompRelDeclaration* *AttributeDeclarations* ;
*DecompRelDeclaration* ::= `mandatory` | `optional` | `alternative to` *FeatureName* | `or to` *FeatureName*
*AttributeDeclarations* ::= ϵ
| *AttributeDec* *AttributeDeclarations*
*AttributeDec* ::= `attribute` *AttributeName* *AssignedValue*
*AssignedValue* ::= *IntegerLiteral* | *RealLiteral* | *BoolLiteral* | *StringLiteral*

*CrossTreeConstraintDeclarations* ::= ϵ
| *CrossTreeConstraintDec* *CrossTreeConstraintDeclarations*
*CrossTreeConstraintDec* ::= `constraint` *FeatureName* *BasicCrossTreeConstraint* *FeatureName* ;

*Commands* ::= *ACommand* | *ACommand* *Commands*
ACommand ::= *AddFeature* | *UpdateFeature* | *RemoveFeature*
| *UpdateMultipleFeatures* | *RemoveMultipleFeatures*
| *AddConstraint* | *UpdateConstraint* | *RemoveConstraint*
| *UpdateMultipleConstraints* | *RemoveMultipleConstraints*

*AddFeature* ::= `add feature` *FeatureName*
`with attributes` *AttributeList*
*WhereClause* ;

*UpdateFeature* ::= `update feature` *FeatureDescription*
`set` *FeatureUpdates*
*WhereClause* ;

*UpdateMultipleFeatures* ::= `updateall feature` *FeatureVar*
`set` *LimitedFeatureUpdates*
*WhereClause* ;

*RemoveFeature* ::= `remove feature` *FeatureDescription*
*WhereClause* ;

*RemoveMultipleFeatures* ::= `removeall feature` *FeatureVar*
*WhereClause* ;

*AddConstraint* ::= `add constraint` *ConstraintDescription*
*WhereClause* ;

*UpdateConstraint* ::= `update constraint` *ConstraintDescription*
`set` *ConstraintUpdates*
*WhereClause* ;

*UpdateMultipleConstraints* ::= `updateall constraint` *ConstraintDescription*
`set` *LimitedConstraintUpdates*
*WhereClause* ;

*RemoveConstraint* ::= `remove constraint` *ConstraintDescription*
*WhereClause* `;`

*RemoveMultipleConstraints* ::= `removeall constraint` *ConstraintDescription*
*WhereClause* `;`

*FeatureDescription* ::= *FeatureName*
| *FeatureVar*

*FeatureNameDescription* ::= *FeatureName*
| *FeatureVar* `.` `_name`

*ConstraintDescription* ::= *FeatureDescription* *BasicCrossTreeConstraint* *FeatureDescription*

*WhereClause* ::= ϵ
| `where` *BooleanExpression*

*AttributeList* ::= `(` *StructuralAttributeAssignments* *AttributeAssignments* `)`

*StructuralAttributeAssignments* ::= *SettingLocation* `,` *SettingDecomposition*
| *SettingDecomposition* `,` *SettingLocation*
*SettingLocation* ::= `_parent =` *FeatureNameDescription*
*SettingDecomposition* ::= `_decomp =` *DecompRelValue*
| `_decomp =` *DecompRelValue* `to` *FeatureDescription*

*AttributeAssignments* ::= ϵ
| `,` *AttrAssign* *AttributeAssignments*
*AttrAssign* ::= *AttributeName* `=` *AttributeValue*
*AttributeValue* ::= `inherited` `:` *FeatureDescription* `.` *AttributeName*
| `numeric` `:` *ArithmeticExpression*
| `boolean` `:` *BooleanExpression*
| `string` `:` *StringLiteral*

*FeatureUpdates* ::= *FeatUpdate* | *FeatUpdate* `,` *FeatureUpdates*
*FeatUpdate* ::= *SettingName*
| *SettingLocation*
| *SettingDecomposition*
| *AttrAssign*
*SettingName* ::= `_name =` *StringLiteral*

*LimitedFeatureUpdates* ::= *LimitedFeatUpdate* | *LimitedFeatUpdate* `,` *LimitedFeatureUpdates*
*LimitedFeatUpdate* ::= *SettingLocation*
| *SettingDecomposition*
| *AttrAssign*

*ConstraintUpdates* ::= *ConstUpdate* | *ConstUpdate* `,` *ConstUpdate* | *ConstUpdate* `,` *ConstUpdate* `,` *ConstUpdate*
*ConstUpdate* ::= `leftfeature` `=` *FeatureNameDescription*
| `constrainttype` `=` *BasicCrossTreeConstraint*
| `rightfeature` `=` *FeatureNameDescription*

*LimitedConstraintUpdates* ::= *ConstUpdate* | *ConstUpdate* `,` *ConstUpdate*

```
DecompRelValue ::= mandatory | optional | alternative | or | FeatureDescription . _decomp

BasicCrossTreeConstraint ::= requires | excludes

ArithmeticExpression ::= ArithmeticOperand
                       | ( ArithmeticExpression )
                       | - ArithmeticExpression
                       | ArithmeticExpression ArithOp ArithmeticExpression
ArithmeticOperand ::= IntegerLiteral
                    | RealLiteral
                    | FeatureDescription . AttributeName
ArithOp ::= + | - | * | / | %

BooleanExpression ::= BooleanOperand
                    | ( BooleanExpression )
                    | not BooleanExpression
                    | BooleanExpression and BooleanExpression
                    | BooleanExpression or BooleanExpression
BooleanOperand ::= BoolLiteral
                 | FeatureDescription . AttributeName
                 | DecompRelTypeCheck
                 | DecompRelIDCheck
                 | BooleanEqualityCheck
                 | StringEqualityCheck
                 | ArithmeticExpression RelOp ArithmeticExpression
RelOp ::= < | <= | > | >= | = | <>

DecompRelTypeCheck ::= DecompRelValue = DecompRelValue
                     | DecompRelValue <> DecompRelValue

DecompRelIDCheck ::= FeatureDescription . _decompID = FeatureDescription . _decompID
                   | FeatureDescription . _decompID <> FeatureDescription . _decompID

BooleanEqualityCheck ::= BoolEqOperand = BoolEqOperand
                       | BoolEqOperand <> BoolEqOperand
BoolEqOperand ::= BoolLiteral
                | FeatureDescription . AttributeName

StringEqualityCheck ::= StringOperand = StringOperand
                      | StringOperand <> StringOperand
StringOperand ::= StringLiteral
                | FeatureDescription . _name
                | FeatureDescription . _parent
                | FeatureDescription . AttributeName


BoolLiteral ::= true | false

IntegerLiteral ::= Sign Digit DigitSeq
Sign           ::= ϵ | + | -
DigitSeq       ::= ϵ | Digit DigitSeq

RealLiteral ::= Sign Digit DigitSeq . Digit DigitSeq
```

*StringLiteral* ::= " *AChar* *StringSeq* "
*AChar* ::= *Letter*
| *Digit*
| *OtherChar*
*StringSeq* ::= ϵ
| *Letter* *StringSeq*
| *Digit* *StringSeq*
| *OtherChar* *StringSeq*

*FeatureName* ::= *StringLiteral*
*ParentName* ::= *StringLiteral*

*FeatureVar* ::= *UpperCase FeatVarSeq*
*FeatVarSeq* ::= ϵ | *AllowedFeatVarChar* *FeatVarSeq*
*AllowedFeatVarChar* ::= *Letter* | *Digit* | '_'

*AttributeName* ::= *LowerCase* *AttrNameSeq*
*AttrNameSeq* ::= ϵ | *AllowedAttrNameChar* *AttrNameSeq*
*AllowedAttrNameChar* ::= *Letter* | *Digit* | '_'

*Letter* ::= *UpperCase* | *LowerCase*
*UpperCase* ::= one of
A B C D E F G H I J K L M N O P Q R S T U V W X Y Z
*LowerCase* ::= one of
a b c d e f g h i j k l m n o p q r s t u v w x y z

*Digit* ::= one of
0 1 2 3 4 5 6 7 8 9

*OtherChar* ::= one of
~ ! @ # $ % ^ & * ( ) _ + [ ] ' / . , - ; :

## Appendix B: Context-free Grammar of the Accepted TVL Subset

*Model* ::= *StringType* *FeatureList*

*StringType* ::= ϵ
| `enum` *string* `in` { *StringList* } `;`

*StringList* ::= *StringLiteral* | *StringLiteral* `,` *StringList*

*FeatureList* ::= *RootFeature* *OtherFeatures*

*RootFeature* ::= `root` *Feature*

*OtherFeatures* ::= ϵ
| *Feature* *OtherFeatures*

*Feature* ::= *ID* { *FeatureBody* }

*FeatureBody* ::= *AttributeList* *Children* *ConstraintList*

*AttributeList* ::= ϵ
| *Attribute* *AttributeList*

*Attribute* ::= `int` *AttrID* `is` *IntegerLiteral* **;**
| `real` *AttrID* `is` *RealLiteral* **;**
| `bool` *AttrID* `is` *BooleanLiteral* **;**
| *string* *AttrID* `is` *StringLiteral* **;**

*Children* ::= ϵ
| *FeatureGroup* *Children*

*FeatureGroup* ::= `group allof` { *SolitaryIDList* }
| `group` *Cardinality* { *IDList* }

*SolitaryIDList* ::= *Optional* *ID*
| *Optional* *ID* **,** *SolitaryIDList*

*Optional* ::= ϵ | `opt`

*Cardinality* ::= `oneof` | `someof`

*IDList* ::= *ID* | *ID* **,** *IDList*

*ConstraintList* ::= ϵ
| *Constraint* *ConstraintList*

*Constraint* ::= *ID* `requires` *ID* **;**
| *ID* `excludes` *ID* **;**

*BooleanLiteral* ::= `true` | `false`

*IntegerLiteral* ::= *Sign* *Digit* *DigitSeq*
*Sign* ::= ϵ | + | -
*DigitSeq* ::= ϵ | *Digit* *DigitSeq*

*RealLiteral* ::= *Sign* *Digit* *DigitSeq* **.** *Digit* *DigitSeq*

*StringLiteral* ::= **"** *AChar* *StringSeq* **"**
*AChar* ::= *Letter*
| *Digit*
| *OtherChar*
*StringSeq* ::= ϵ
| *Letter* *StringSeq*
| *Digit* *StringSeq*
| *OtherChar* *StringSeq*

*ID* ::= *Letter* *IDSeq*
*IDSeq* ::= ϵ | *AllowedIDChar* *IDSeq*
*AllowedIDChar* ::= *Letter* | *Digit* | '_'

*AttrID* ::= *LowerCase* *IDSeq*

*Letter* ::= *UpperCase* | *LowerCase*
*UpperCase* ::= one of
A B C D E F G H I J K L M N O P Q R S T U V W X Y Z
*LowerCase* ::= one of
a b c d e f g h i j k l m n o p q r s t u v w x y z

*Digit* ::= one of
0 1 2 3 4 5 6 7 8 9

*OtherChar* ::= one of
~ ! @ # $ % ^ & * ( ) _ + [ ] ' / . , - ; :